\documentclass[a4paper,11pt]{article}
\pdfoutput=1 

\usepackage{jcappub} 

\usepackage[T1]{fontenc} 
\usepackage{caption}

\newcommand{\beq}{\begin{equation}}
\newcommand{\eeq}{\end{equation}}
\newcommand{\bea}{\begin{eqnarray}}
\newcommand{\eea}{\end{eqnarray}}
\newcommand{\beqn}{\begin{equation*}}
\newcommand{\eeqn}{\end{equation*}}
\newcommand{\bean}{\begin{eqnarray*}}
\newcommand{\eean}{\end{eqnarray*}}

\newcommand*{\cref}[1]{Chapter~\ref{#1}}

\title{\boldmath On the $\alpha$-attractor T-models}

\author[a]{Gabriel Germ\'an}

\affiliation[a]{Instituto de Ciencias F\'{\i}sicas, Universidad Nacional Aut\'onoma de M\'exico, Av. Universidad s/n, Cuernavaca, Morelos 62210, Mexico}

\emailAdd{gabriel@icf.unam.mx}

\abstract{We carry out a fully analytical study of the phenomenology of $\alpha$-attractor T-models defined by the potential $V = V_0\tanh^ p\left(\lambda \phi/M_{pl}\right)$. We obtain expressions for the number of e-folds during inflation $N_{ke}$ in terms of the scalar spectral index $n_s$ and independently in terms of the tensor-to-scalar ratio $r$. From these expressions we obtain exact solutions for both $n_s$ and $r$ in terms of  $N_{ke}$ along with their expansions for large $N_{ke}$, in full agreement with known expressions. Eliminating the parameter $\lambda$ from the model in terms of $n_s$ and $r$ we can obtain exact solutions for $r$ in terms of $n_s$ and $N_{ke}$ which allows us to reproduce, in particular, numerical solutions presented by the Planck Collaboration for the monomial potentials. We explicitly show how these solutions are contained in the solutions for the $\alpha$-attractors and are also the end points of these. Finally, by also eliminating the global scale $V_0$ in terms of the observables $n_s$ and $r$ we show how in the appropriate limit the $\alpha$-attractor potential exactly reduces to the monomials potential. We also briefly show that for $\alpha$-attractor E-models, which generalize the Starobinsky potential in the Einstein frame, a similar transition occurs.}

\begin{document}
\maketitle
\flushbottom

\section {\bf Introduction}\label{INT}

In recent years alpha attractor-type models have captivated considerable attention mainly due to the fact that they have a fairly well-understood origin in conformal and super-conformal field theories as well as their close relationship with supergravity theories \cite{Kallosh:2013hoa}-\cite{Kallosh:2016sej}, the fact that they connect with well known monomial potentials and most importantly, because they have a phenomenology \cite{Kallosh:2013yoa} that is fully consistent with reported results  by e.g., the Planck Collaboration \cite{Akrami:2018odb}.
Important properties of $\alpha$-attractors have been extensively discussed in the literature (for a sample of articles in the subject see e.g., \cite{Odintsov:2016vzz}-\cite{Akrami:2020zxw} and references therein). The resulting class of potentials generalized from the simplest monomials is of the form 
\begin{equation}
V = V_0\tanh^ p\left(\lambda \frac{\phi}{M_{pl}}\right),  \quad\quad\quad (T-models)
\label{pot}
\end{equation}
and 
\begin{equation}
V = V_0\left(1-e^{-\lambda\sqrt{\frac{2}{3}}\frac{\phi}{M_{pl} }}\right)^p.  \quad\quad\quad (E-models)
\label{potEmodel}
\end{equation}
Connecting with the original notation $\lambda = 1/ \sqrt{6\alpha}$ for T-models or  $\lambda = 1/ \sqrt{\alpha}$ for E-models thus, $``\alpha"$-attractors. These potentials can be considered on its own as phenomenological potentials for inflation and as such have also been studied, mainly numerically.  The main purpose of this work is to provide a fully $analytical$ treatment, which can be considered as complementary to phenomenological results briefly presented in the literature \cite{Kallosh:2013yoa}, of the most important properties of the T-type $\alpha$-attractors defined by the potential \eqref{pot}.

The organization of the article is as follows: In Section \ref{THE} we discuss in detail the end of slow-roll either with the condition $\eta = -1$ or with $\epsilon = 1$ where $\eta$ and $\epsilon$ are the usual slow-roll parameters given by \cite{Lyth:1998xn}
\begin{equation}
\epsilon \equiv \frac{M_{pl}^{2}}{2}\left( \frac{V^{\prime }}{V }\right) ^{2},\quad\quad
\eta \equiv M_{pl}^{2}\frac{V^{\prime \prime }}{V},
\label{Spa}
\end{equation}
$M_{pl}$ is the reduced Planck mass $M_{pl}=2.44\times 10^{18} \,\mathrm{GeV}$ which we set equal to 1 in most of what follows and primes on $V$ denote derivatives with respect to the inflaton field $\phi$. We provide expressions for the number of e-folds $N_{ke}$ from the time scales left the horizon at wavenumber mode $k$ corresponding to $\phi_k$ to end of inflation at $\phi_e$. The expression for $N_{ke}$ when $\phi_k=\phi_k(n_s)$ is exclusively dependent on the spectral index $n_s$, the model characterized by $p$ and the $\lambda$ parameter appearing in the potential.  This is done by obtaining $\phi_k$ from the equation \cite{Lyth:1998xn}
\begin{equation}
n_{s} =1+2\eta -6\epsilon ,  
\label{Ins} \\
\end{equation}
written in the form $\delta_{n_s}+2\eta-6\epsilon=0$, where $\delta_{n_s}$ is defined as $\delta_{n_s}\equiv 1-n_s$. 
When $\phi_k=\phi_k(r)$ we obtain the number of e-folds during inflation $N_{ke}$ exclusively dependent on the tensor-to-scalar ratio $r$, $p$ and $\lambda$ by solving for $\phi_k$ from the equation
\begin{equation}
n_{t} =-2\epsilon = -\frac{r}{8} ,
\label{Int} 
\end{equation}
written in the form $r=16\epsilon.$ In this way we can obtain exact solutions for $n_s (N_{ke}, \lambda, p)$ and for $r(N_{ke}, \lambda, p)$  as well as their asymptotic behavior in various situations, mainly for large number of e-folds $N_{ke}$. 
In Section \ref{REM} we also obtain an expression for $N_{ke}$ but this time in terms of $r$, $n_s$ and $p$ by eliminating the parameter $\lambda$  in terms of $n_s$ and $r$. In this case we can write any quantity of interest exclusively in terms of the observables $n_s$ and $r$, the quantities  so written will keep tightening as more precise determination of the observables is achieved. 

Using the allowed range of $\lambda$ we can deduce the range for $r$. In particular we find solutions for $r(n_s,N_{ke}, p)$ that allow us to study the $n_s$-$r$ plane for various values of $N_{ke}$ and $p$. Using an approximation for $r$ in the large $N_{ke}$ limit we find bounds for $N_{ke}$ as well as a lower bound for the parameter $p$.
We explicitly show how the predictions for $r(n_s, p)$ of the monomials potential  $V_{mon}=\frac{1}{2}m^{4-p}\phi^p$ are exactly contained in the solutions for $r$ of the $\alpha$-attractor models and should also be the ending points of the curves $r(n_s,N_{ke},p)$ \cite{Kallosh:2013yoa}. To fully clarify this phenomenon we also determine the global scale $V_0$ in terms of $n_s$ and $r$ through the equation
\begin{equation}
A_s(k) =\frac{1}{24\pi ^{2}} \frac{V}{M^4\epsilon}.
\label{IA} 
\end{equation}
Thus, a direct study of the potential \eqref{pot} with $V_0$ and $\lambda$ eliminated in terms of $n_s$ and $r$ shows how it transitions exactly to the monomials potential in the limiting case in which the relationship between the observables $r$ and $n_s$ of the monomials is fulfilled and we  illustrate graphically this phenomenon in Fig.\,\ref{pottanmon}. We show briefly that this is also the case for the type of E-models \eqref{potEmodel} that can be considered as generalizations of the Starobinsky model in the Einstein frame (see Fig.\,\ref{potstaro}). Finally, Section \ref{CON} contains our conclusions on the main points discussed in the article.

\section {\bf The $\alpha$-attractor T-models}\label{THE} 

\begin{figure}[tb]
\begin{center}
\includegraphics[width=10cm]{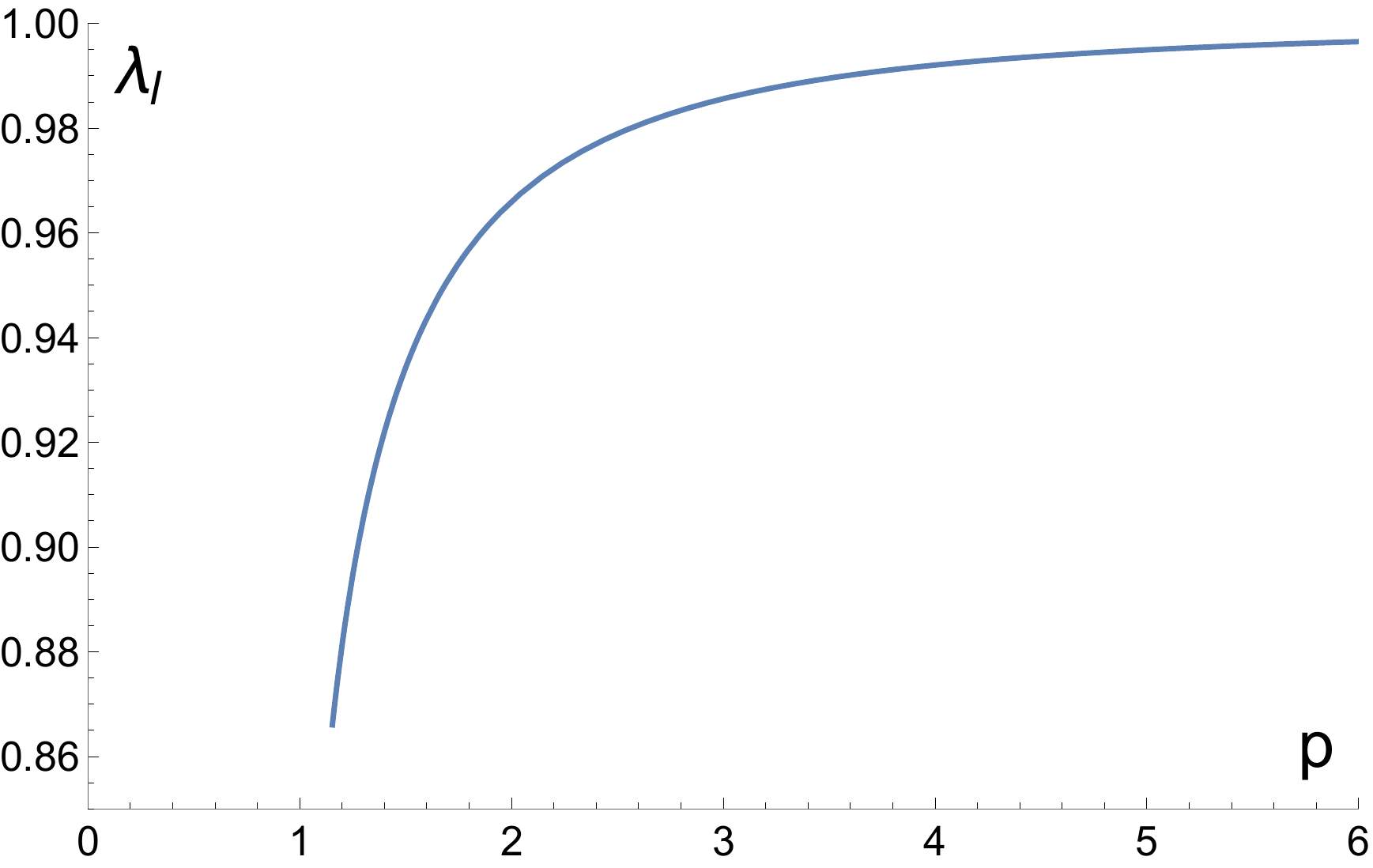}
\caption{\small The figure shows the limiting value of $\lambda$ as a function of $p$, given by Eq.~\eqref{Lmin}, which separates solutions where the end of slow-roll is given by the condition $\eta=-1$ from those where inflation is terminated by $\epsilon=1$. For $p=2$, $\lambda_l   \approx 0.9659$.}
\label{Ll}
\end{center}
\end{figure}

\begin{figure}[t!]
\begin{center}
\includegraphics[trim = 0mm  0mm 1mm 1mm, clip, width=7.5cm, height=5.5cm]{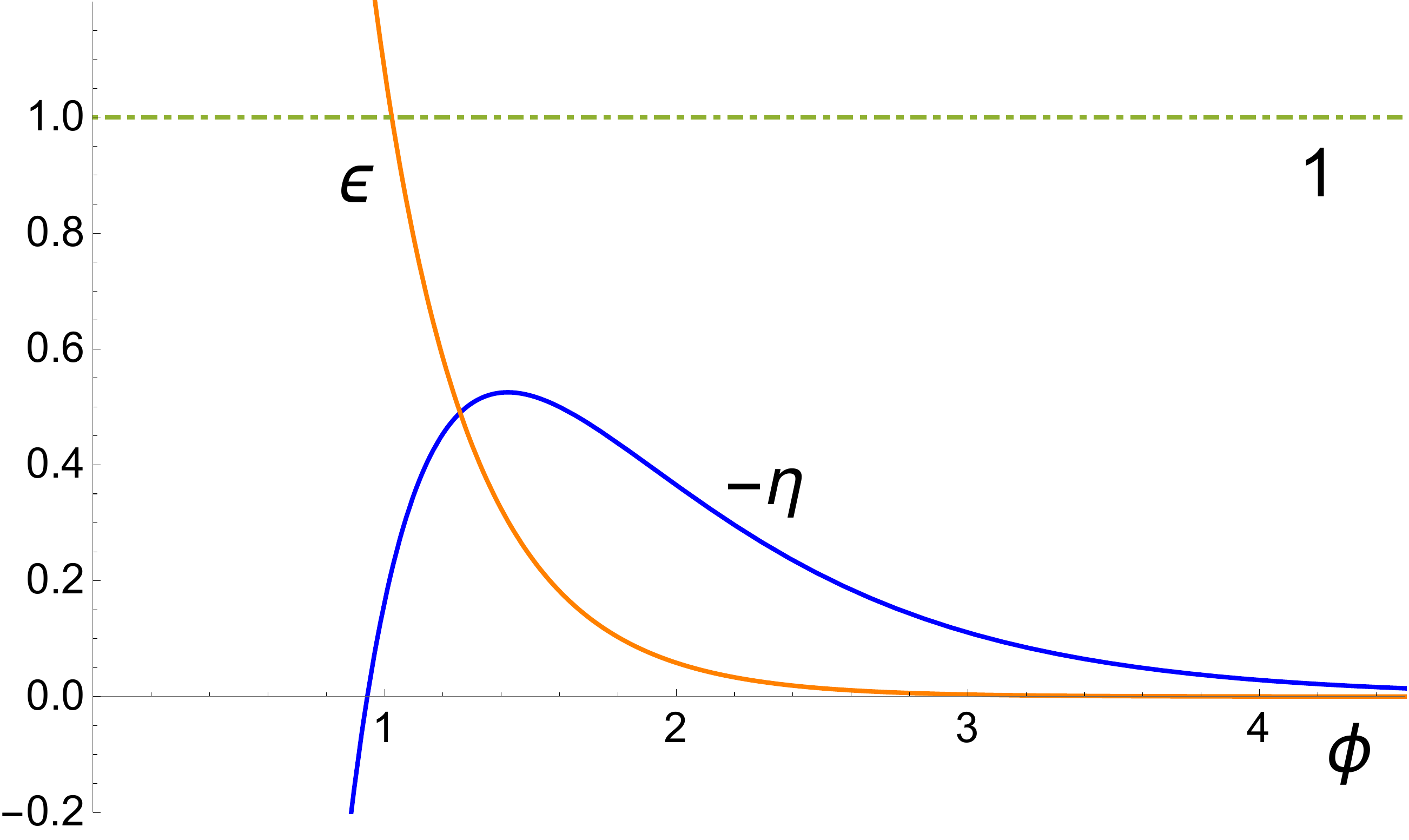}
\includegraphics[trim = 0mm  0mm 1mm 1mm, clip, width=7.5cm, height=5.5cm]{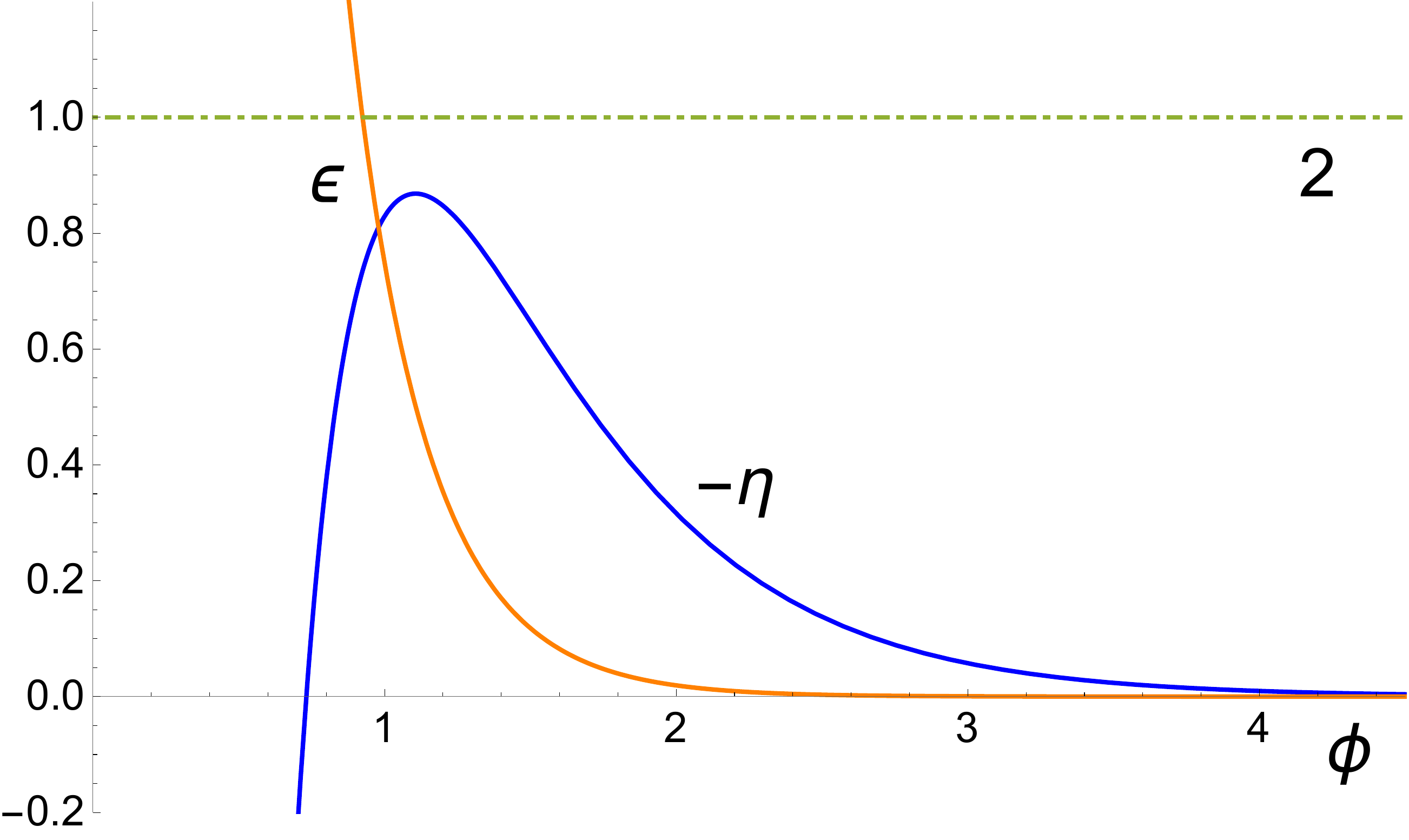}
\includegraphics[trim = 0mm  0mm 1mm 1mm, clip, width=7.5cm, height=5.5cm]{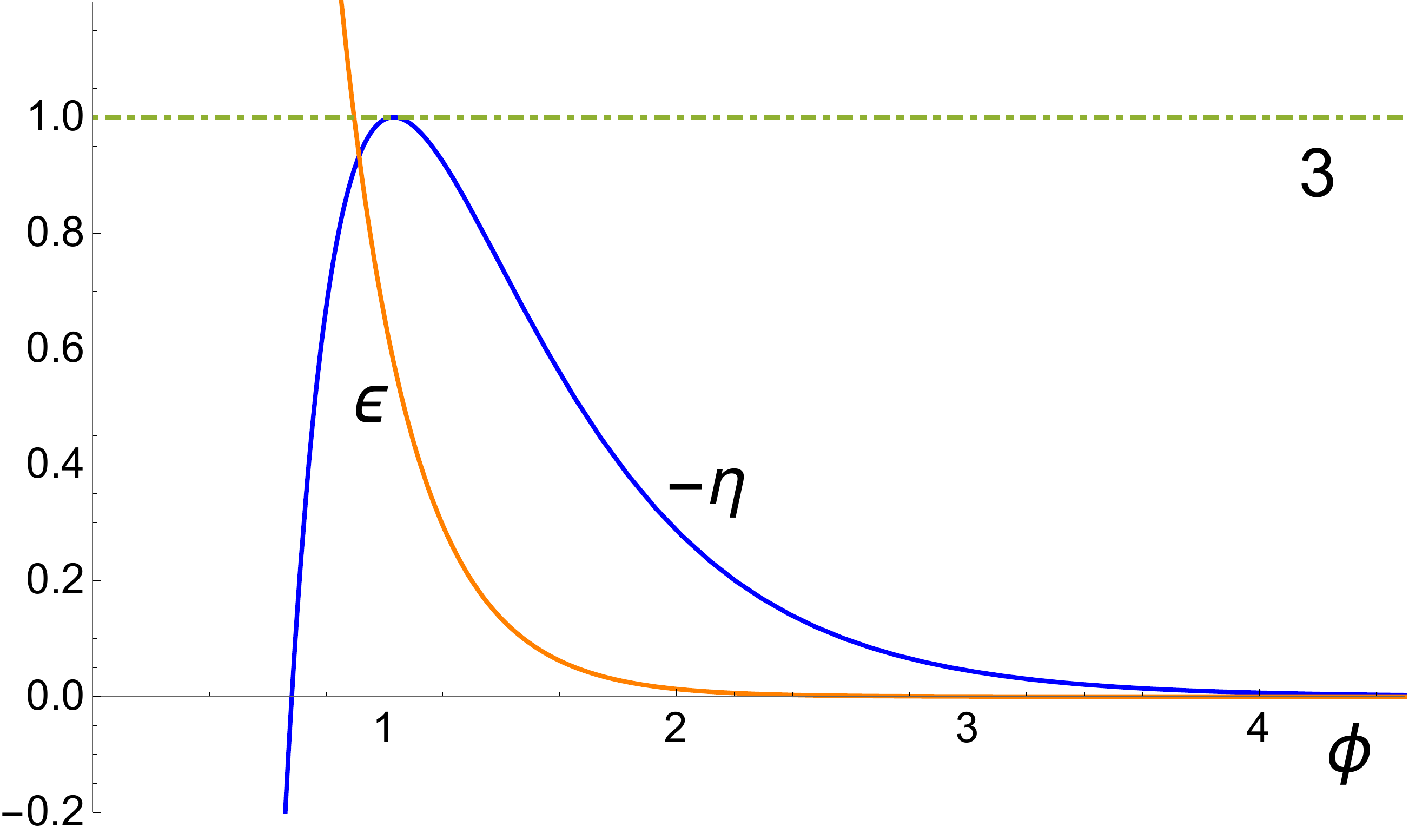}
\includegraphics[trim = 0mm  0mm 1mm 1mm, clip, width=7.5cm, height=5.5cm]{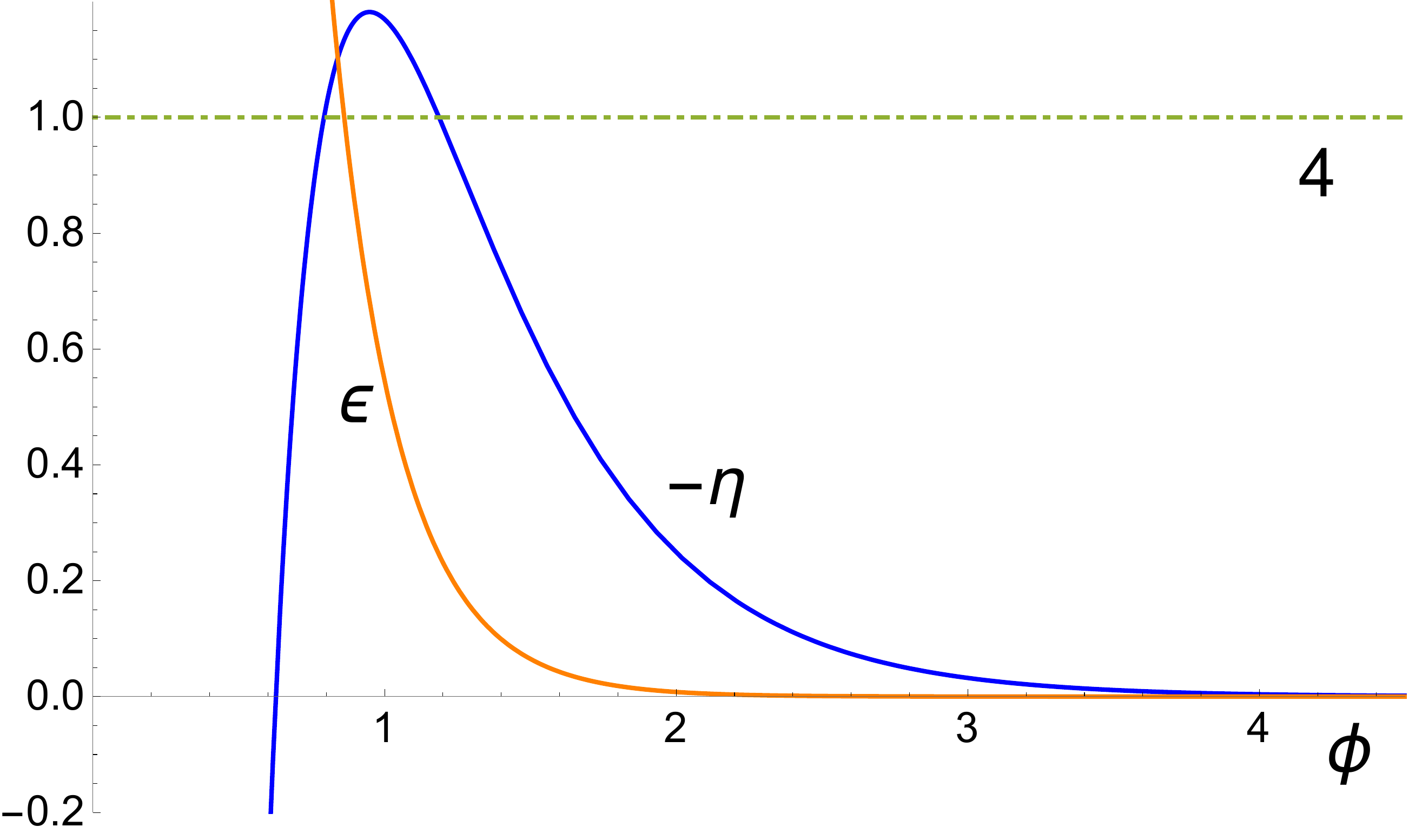}
\end{center}
\caption{Plot of the slow-roll parameters $\epsilon$ and $-\eta$ as functions of the inflaton $\phi$ for values of $\lambda$ smaller than $\lambda_l$ (panels 1 and 2) for $\lambda=\lambda_l$ (panel 3) and for $\lambda>\lambda_l$. We see in panel 3 how slow-roll is terminated by the saturation of the condition $\eta=-1$ to briefly reset slow-roll and finally ending inflation by the condition $\epsilon=1$ (see Fig.\,\ref{srint} below).}
\label{li}
\end{figure}

The $\alpha$-attractor T-models generalized from the simplest monomials are defined by the potential given by Eq.~\eqref{pot}, where $p$ is a positive number which distinguishes among the models and $\lambda$ is a parameter.  When necessary we can understand $\phi$ in Eq.~\eqref{pot} as its absolute value to guarantee a bounded potential from below. We do not write it explicitly anywhere because we will be working in the positive region of $\phi$ always unless stated otherwise.
An expression for $\phi_k$, the inflaton at horizon crossing, is obtained  by solving Eq.~\eqref{Ins}, $\delta_{n_s}+2\eta-6\epsilon=0$, with the result
\begin{equation}
\cosh^2\left(\lambda\frac{\phi_k}{M_{pl}}\right)=\frac{1}{2\delta_{n_s}}\left(\delta_{n_s}+4p\lambda^2+\sqrt{\delta_{n_s}^2+4p^2\lambda^2\delta_{n_s}+16p^2\lambda^4}\right)\;,
\label{fik1a}
\end{equation}
where $\delta_{n_s}$ is defined as $\delta_{n_s}\equiv 1-n_s$. For large $\lambda$ the end of slow-roll is given by the solution to the equation $\eta=-1$ while for small $\lambda$ by the condition $\epsilon=1$. The value of $\phi$ which solves the condition $\eta=-1$ is $\phi_{e\eta}$ and it is given by
\begin{equation}
\cosh^2\left(\lambda\frac{\phi_{e\eta}}{M_{pl}}\right)=\frac{1}{2}\left(1+2p\lambda^2+\sqrt{1+4p^2\lambda^2(\lambda^2-1)}\right)\;,
\label{fie1a}
\end{equation}
while the solution for the $\epsilon=1$ case is 
\begin{equation}
\cosh^2\left(\lambda\frac{\phi_{e\epsilon}}{M_{pl}}\right)=\frac{1}{2}\left(1+\sqrt{1+2p^2\lambda^2}\right)\;.
\label{fie2a}
\end{equation}
The solution Eq.~\eqref{fie1a} makes sense for $1+4p^2\lambda^2(\lambda^2-1)\geq 0$. Thus, the value of $\lambda$ which separates one $\phi_e$-solution from the other is obtained by solving $1+4p^2\lambda^2(\lambda^2-1)= 0$ with the result (see Fig.\,\ref{Ll})
\begin{equation}
\lambda_l=\frac{1}{\sqrt{2}}\sqrt{1+(p^2-1)^{1/2}/p}\;,
\label{Lmin}
\end{equation}
whenever $p\geq 2/\sqrt{3}$. The value of $\lambda_l$ signals the minimum value  $\lambda$ can have (for a given $p)$ when  $-\eta=1$ (see Fig.\,\ref{li}, in particular panel number 3). Thus, for $\lambda<\lambda_l$ the end of inflation is dictated by the condition $\epsilon=1$ while the case $\lambda\geq \lambda_l$ requires solving the equation $-\eta=1$ for the end of slow-roll with the solution given  above. For $p=2$, $\lambda_l   \approx 0.9659$.

The number of e-folds from the time scales the order of the pivot scale left the horizon during inflation at $a_k$ to the end of inflation at $a_e$ is given by 
\begin{equation}
N_{ke} = -\frac{1}{M_{pl}^2}\int_{\phi_k}^{\phi_e}\frac{V}{V'}d\phi=\frac{1}{2p\lambda^2}\left(\cosh^2\left(\lambda\frac{\phi_k}{M_{pl}}\right)-\cosh^2\left(\lambda\frac{\phi_e}{M_{pl}}\right)\right)\;.
\label{Nke}
\end{equation}
We can write $N_{ke}$ in the form $N_{ke}=N_{k}-N_{e}$ where  
\begin{equation}
N_{k} \equiv\frac{1}{2p\lambda^2}\cosh^2\left(\lambda\frac{\phi_k}{M_{pl}}\right)\;,
\label{Nk}
\end{equation}
and 
\begin{equation}
N_{e} \equiv\frac{1}{2p\lambda^2}\cosh^2\left(\lambda\frac{\phi_e}{M_{pl}}\right)\;.
\label{Ne}
\end{equation}
\noindent
\subsection {\bf Slow-roll interruptus }\label{INF} 
\begin{figure}[tb]
\begin{center}
\includegraphics[trim = 0mm  0mm 1mm 1mm, clip, width=7.5cm, height=5.5cm]{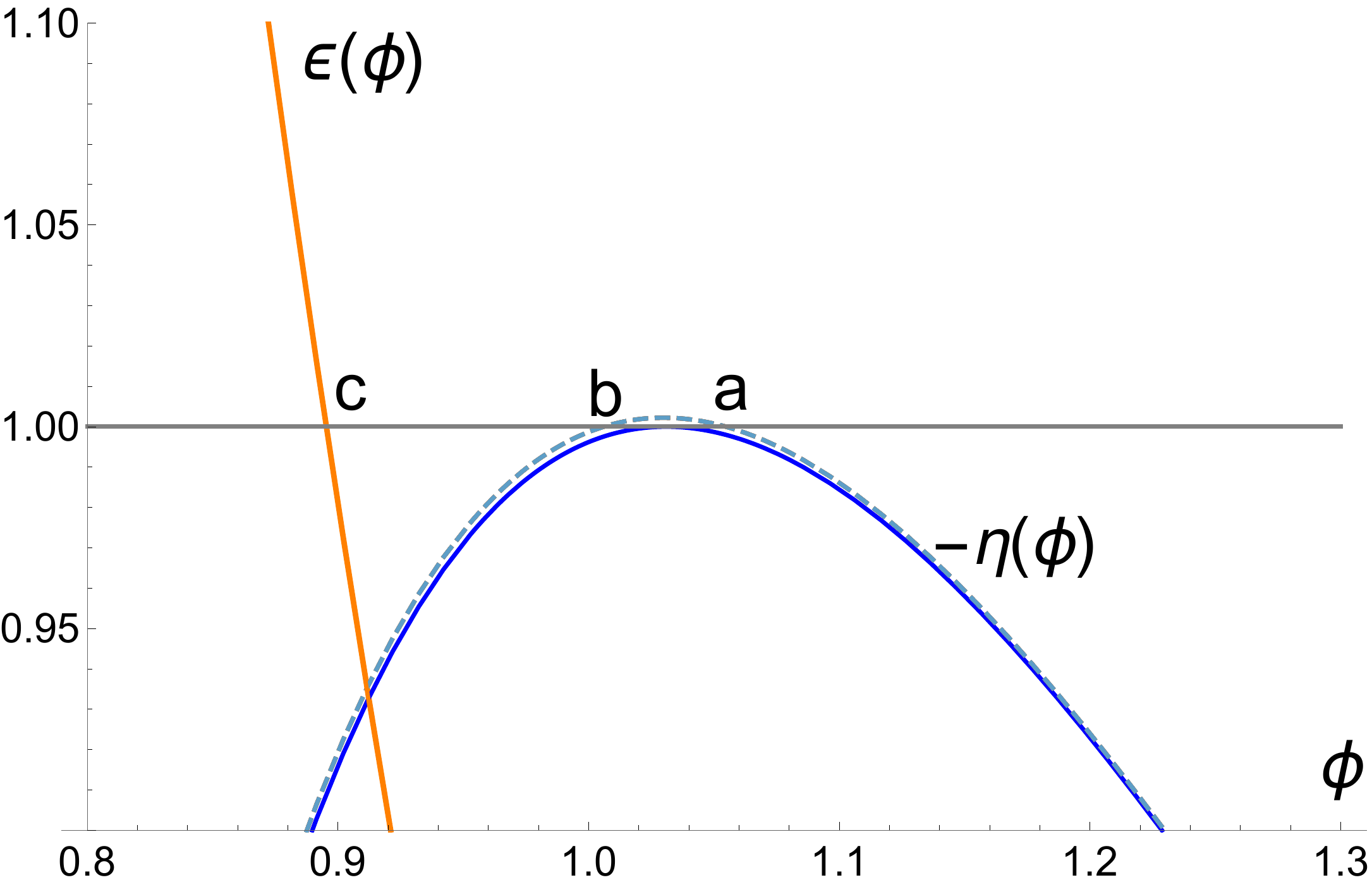}
\includegraphics[trim = 0mm  0mm 1mm 1mm, clip, width=7.5cm, height=5.5cm]{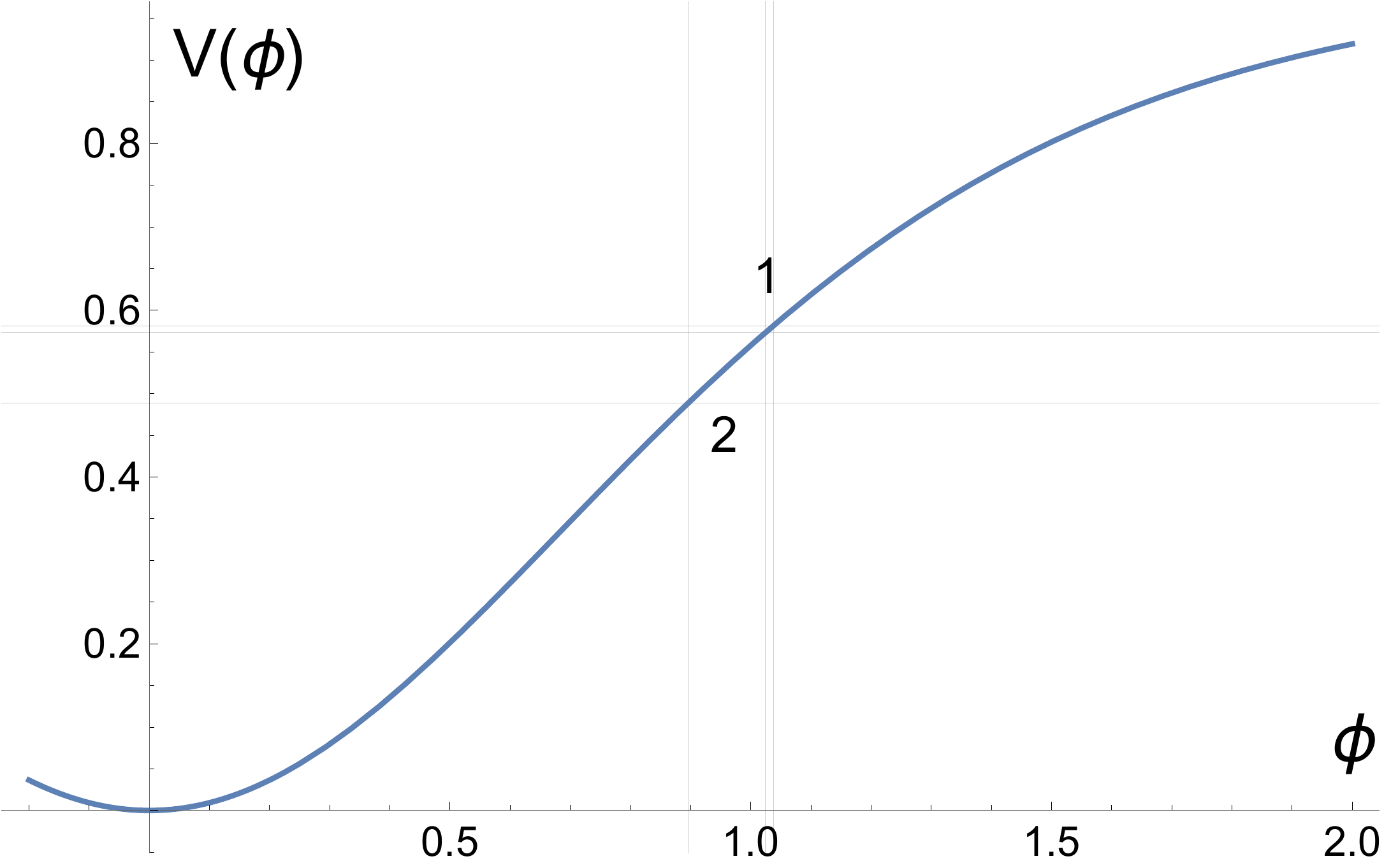}
\caption{\small The l.h.s figure shows $-\eta$ for $\lambda=\lambda_l$ (bottom curve) (see Fig.\,\ref{li} panel 3) and for a slightly larger value of $\lambda$. We see how the upper curve reaches the condition $-\eta=1$ at the point $a$ to reenter again at $b$ the region  where both $-\eta$ and $\epsilon$ are less than one reseting slow-roll to finally end inflation when $\epsilon(\phi)=1$ at the point $c$. We also show the potential (for $\lambda\geq \lambda_l$) in the r.h.s figure where the very small box under the number 1 shows the region where slow-roll has ended and in the square above 2 the region where slow-roll is restored to eventually end inflation permanently in the l.h.s. corner of the large square corresponding to $\epsilon=1$. 
}
\label{srint}
\end{center}
\end{figure}

An interesting phenomenon occurs near the end of inflation and we illustrate it below for the case $p=2$. By looking at the third panel in Fig.\,\ref{li} (in Fig.\,\ref{li} we plot $-\eta$) we see that the curve $-\eta$ just touches the horizontal line at 1 for $\lambda=\lambda_l$. This means that the condition $\eta=-1$ has been marginally satisfied signaling the end of slow-roll. In the Fig.\,\ref{srint} this detail is amplified together with a second curve for $-\eta$ corresponding to a value slightly greater than $\lambda_l$ ($\lambda=0.966$). We can see that at the point $a=(1.0374,1)$ the condition $-\eta=1$ is saturated but the curve enters again the region
 $-\eta<1$ at $b=(1.0245,1)$ when $\epsilon$ is still less than 1 i.e., slow-roll resets to eventually end inflation at $c=(0.8956,1)$ when $\epsilon$ is finally equal to one.

In Fig.\,\ref{srint} we also show the potential for $\lambda\geq \lambda_l$ during this phenomenon: the small box under the number 1 in Fig.\,\ref{srint} shows the region during which slow-roll has ended and in the square above 2 the region where slow-roll is restored to eventually end permanently in the l.h.s. corner of the square corresponding to $\epsilon=1$. The total number of e-folds between points $b$ and $c$ is $0.104$ corresponding to an $11\%$ increase of the size of the universe in $b$. For slightly higher values of $\lambda$ slow-roll will be interrupted for a longer period. While the number of e-folds after slow-roll is restored is negligible, the phenomenon itself is interesting and worth reporting.
\subsection {\bf The case  $\lambda\geq \lambda_l$}\label{Larger}

The equations for $N_k$ and $N_e$,  Eqs.~\eqref{Nk} and \eqref{Ne} respectively, are obtained by substituting $\phi_k$ and  $\phi_{e\eta}$ from \eqref{fik1a} and \eqref{fie1a}. Together with the large $\lambda$ expansion they are given by
\begin{equation}
N_{k} =\frac{\delta_{n_s}+4p\lambda^2+\sqrt{\delta_{n_s}^2+4p^2\lambda^2\delta_{n_s}+16p^2\lambda^4}}{4p\lambda^2\delta_{n_s}}=\frac{2}{\delta_{n_s}}+\frac{p+2}{8p\lambda^2}+\cdot\cdot\cdot\;,
\label{Nkapplarge}
\end{equation}
and 
\begin{equation}
N_{e} =\frac{1+2p\lambda^2+\sqrt{1+4p^2\lambda^2(\lambda^2-1)}}{4p\lambda^2}=1-\frac{p-1}{4p\lambda^2}+\cdot\cdot\cdot\;.
\label{Neapplarge}
\end{equation}
We see that both terms are $\lambda$-independent for infinitely large $\lambda$,  the term associated with  the end of slow-roll is negligible contributing with less than 1 e-fold. We will see in Subsection \ref{Smaller}
that when $\lambda$ is small this is not the case and $N_e$ is important.

\begin{figure}[tb]
\begin{center}
\includegraphics[trim = 0mm  0mm 1mm 1mm, clip, width=7.5cm, height=5.5cm]{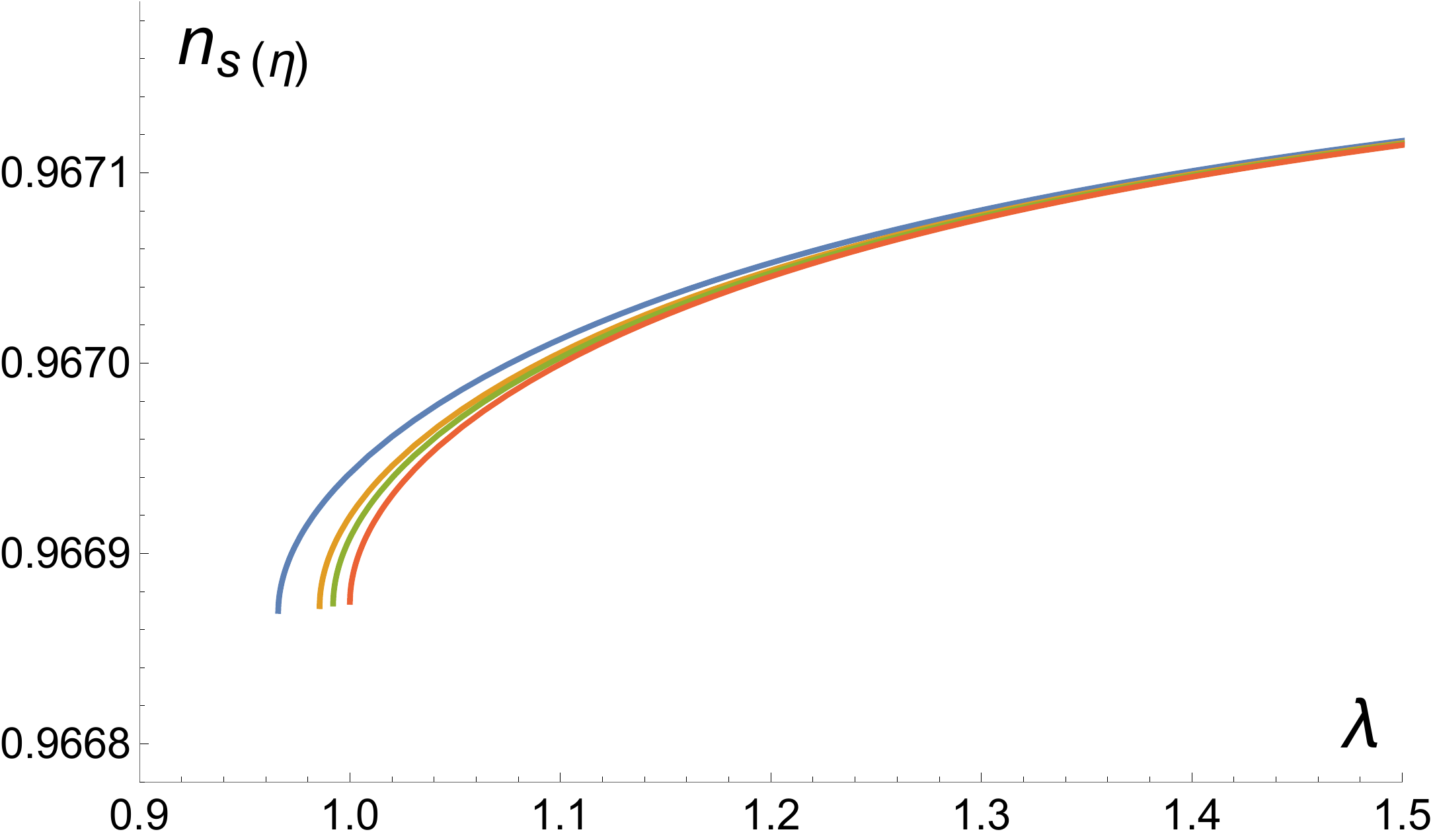}
\includegraphics[trim = 0mm  0mm 1mm 1mm, clip, width=7.5cm, height=5.5cm]{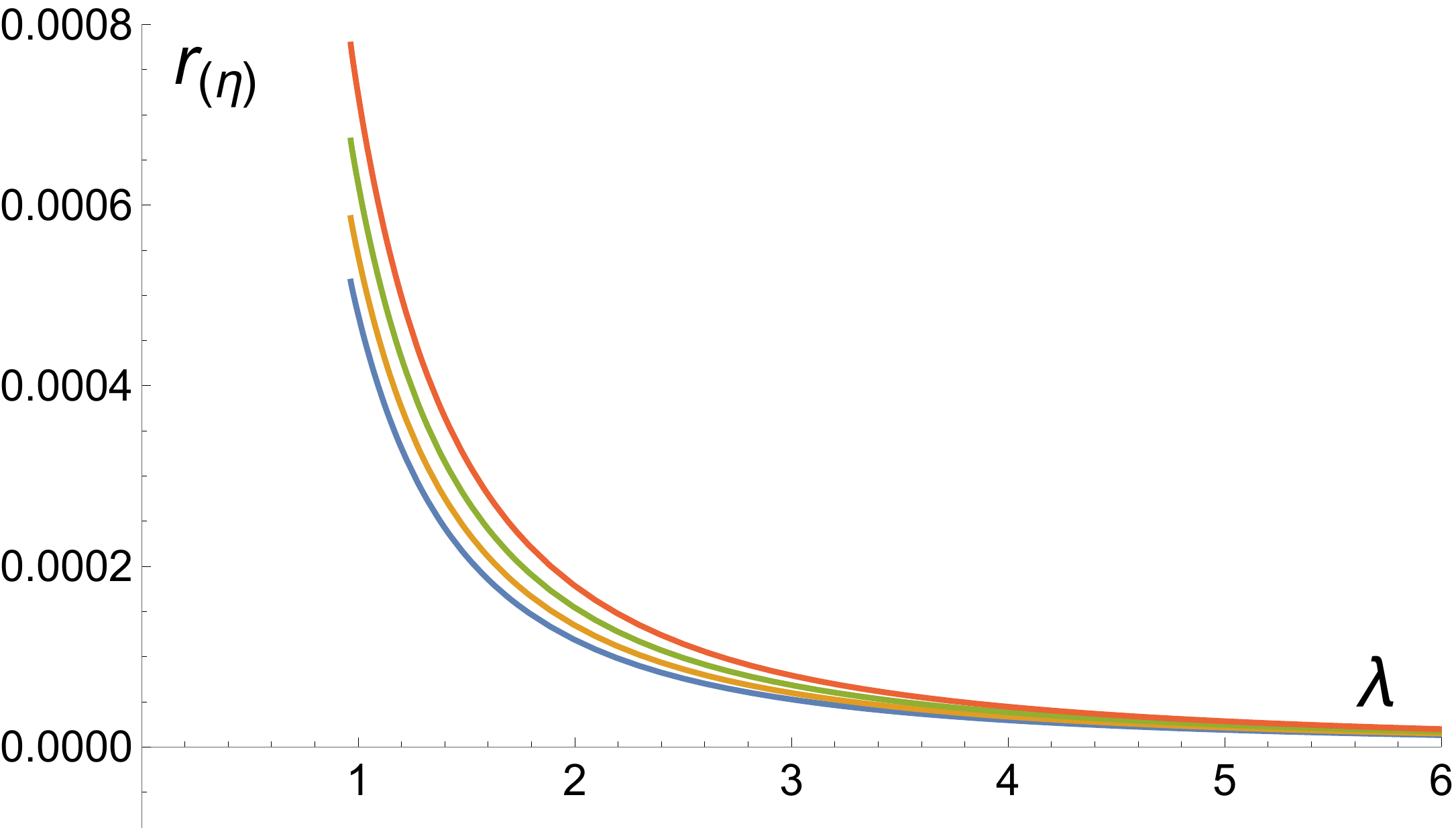}
\caption{\small The l.h.s. figure illustrates the scalar spectral index $n_s$ as a function of $\lambda$ as given by Eq.~\eqref{ns1a} for $N_{ke}=60$ and (from left to right) $p=2,3,4,10^6$ while the r.h.s. figure shows $r_{(\eta)}(\lambda)$ given by Eq.~\eqref{r1b} for $p=2$ and (from left to right) $N_{ke}=64,60,56,52$, both figures with $\lambda>\lambda_l$. There is no appreciable change in $r_{(\eta)}$ when plotted for various values of $p$. The subindex $\eta$ is intended to emphasize that the solution has been obtained under the condition $\eta=-1$ for ending slow-roll.
}
\label{nseta}
\end{center}
\end{figure}

Using Eqs.~\eqref{Nkapplarge} and \eqref{Neapplarge}, the number of e-folds $N_{ke}$ can be written in terms of the observable $n_s$ as follows
\begin{equation}
N_{ke\eta}(n_s,\lambda,p) = \frac{2p(2-\delta_{n_s})\lambda^2+\sqrt{\delta_{n_s}^2+4p^2\lambda^2\delta_{n_s}+16p^2\lambda^4}-\delta_{n_s}R_1}{4p\lambda^2\delta_{n_s}}\;,
\label{Nke1a}
\end{equation}
where $R_1=\sqrt{1+4p^2\lambda^2(\lambda^2-1)}$ and the subindex $\eta$ appears here to remind us that $N_{ke}$ is obtained from the solution to the $\eta=-1$ condition but then it is dropped (as well as the argument) from the following expressions. For large $\lambda$, $N_{ke}$ has the following expansion
\begin{equation}
N_{ke}=\frac{2-\delta_{n_s}}{\delta_{n_s}}+\frac{3}{8\lambda^2}-\frac{8-4\delta_{n_s}-p^2(8-\delta_{n_s})}{128p^2\lambda^4}+\cdot\cdot\cdot\;.
\label{Nke1aexp}
\end{equation}
Thus, $N_{ke}$ is $p$-independent for infinitely large $\lambda$. We can solve Eq.~\eqref{Nke1a} for $n_s$ in terms of $N_{ke}$ (see Fig.\,\ref{nseta})
\begin{equation}
n_s=\frac{1+4p^2\lambda^2-(R_1-2(3-2N_{ke})p\lambda^2)(R_1+2(1+2N_{ke})p\lambda^2)}{1-(R_1+2(1+2N_{ke})p\lambda^2)^2}\;.
\label{ns1a}
\end{equation}
Thus, the leading order term in the large-$N_{ke}$ expansion of the scalar spectral index $n_s$ is also p-independent
\begin{equation}
n_s=1-\frac{2}{N_{ke}}+\left(1+\frac{2R_1-p}{4p\lambda^2}\right)\frac{1}{N_{ke}^2}+\cdot\cdot\cdot\;.,
\label{ns1aexp}
\end{equation}
for the mean value $n_s=0.9649$ \cite{Akrami:2018odb} Eq.~\eqref{ns1aexp} implies $N_{ke}\approx 57$.

One can notice that $N_{ke}$ as given by Eq.~\eqref{Nke1a} above depends only on the observable $n_s$, this is so because $\phi_k$ was obtained by solving the equation \eqref{Ins} written in the form $\delta_{n_s}+2\eta-6\epsilon=0$. We could also obtain $\phi_k$ by solving Eq.~\eqref{Int} or $r=16\epsilon$ and write an equivalent expression for $N_{ke}$ involving only the observable $r$ with the result
\begin{equation}
\cosh^2\left(\lambda\frac{\phi_k}{M_{pl}}\right)=\frac{1}{2}\left( 1+\sqrt{1+\frac{32p^2\lambda^2}{r}}\right)\;,
\label{fik1b}
\end{equation}
which together with Eq.~\eqref{fie1a} implies that the number of e-folds can now be written as
\begin{equation}
N_{ke\eta}(r,\lambda,p) = \frac{-2p\lambda^2+\sqrt{1+\frac{32p^2\lambda^2}{r}}-R_1}{4p\lambda^2}\;.
\label{Nke1b}
\end{equation}
Solving for $r$
\begin{equation}
r=\frac{8p}{p(-1+2(1+2N_{ke}(1+N_{ke}))\lambda^2)+(1+2N_{ke}R_1)}\;,
\label{r1b}
\end{equation}
as shown by Fig.\,\ref{nseta} for the $p=2$ case and for various values of $N_{ke}$.
In the large-$N_{ke}$ limit $r$ is given by
\begin{equation}
r=\frac{2}{\lambda^2 N_{ke}^2}
-\frac{2p\lambda^2+R_1}{p\lambda^4N_{ke}^3}+\cdot\cdot\cdot\;.
\label{r1bexp}
\end{equation} 
\subsection {\bf The case  $\lambda < \lambda_l$}\label{Smaller}

The end of inflation is given by the solution to the condition $\epsilon=1$
\begin{equation}
\cosh^2\left(\lambda\frac{\phi_{e\epsilon}}{M_{pl}}\right)=\frac{1}{2}\left(1+\sqrt{1+2p^2\lambda^2}\right)\;,
\label{fie2arepetida}
\end{equation}
while $\phi_k$ in terms of $n_s$ is given, as before, by Eq.~\eqref{fik1a}.

The equations for $N_k$ and $N_e$, Eqs.~\eqref{Nk} and \eqref{Ne} respectively, and their small $\lambda$ expansion, are given by
\begin{equation}
N_{k} =\frac{\delta_{n_s}+4p\lambda^2+\sqrt{\delta_{n_s}^2+4p^2\lambda^2\delta_{n_s}+16p^2\lambda^4}}{4p\lambda^2\delta_{n_s}}=\frac{1}{2p\lambda^2}+\frac{p+2}{2\delta_{n_s}}-\frac{p(p^2-4)\lambda^2}{2\delta_{n_s}^2}+\cdot\cdot\cdot\;,
\label{Nkappsmall}
\end{equation}
and 
\begin{equation}
N_{e} =\frac{1+\sqrt{1+2p^2\lambda^2}}{4p\lambda^2}=\frac{1}{2p\lambda^2}+\frac{p}{4}-\frac{p^3\lambda^2}{8}+\cdot\cdot\cdot\;.
\label{Neappsmall}
\end{equation}

For small $\lambda$ the first term in the expansion of $N_k$ grows large thus, (contrary to the large $\lambda$ case where $N_{e}$ is less than 1) here the end of inflation is important and necessary to cancel the leading term in $N_k$ so that the number of e-folds of inflation goes like $N_{ke}\approx (p+2)/2\delta_{n_s}$, to first approximation.

\begin{figure}[tb]
\begin{center}
\includegraphics[trim = 0mm  0mm 1mm 1mm, clip, width=7.5cm, height=5.5cm]{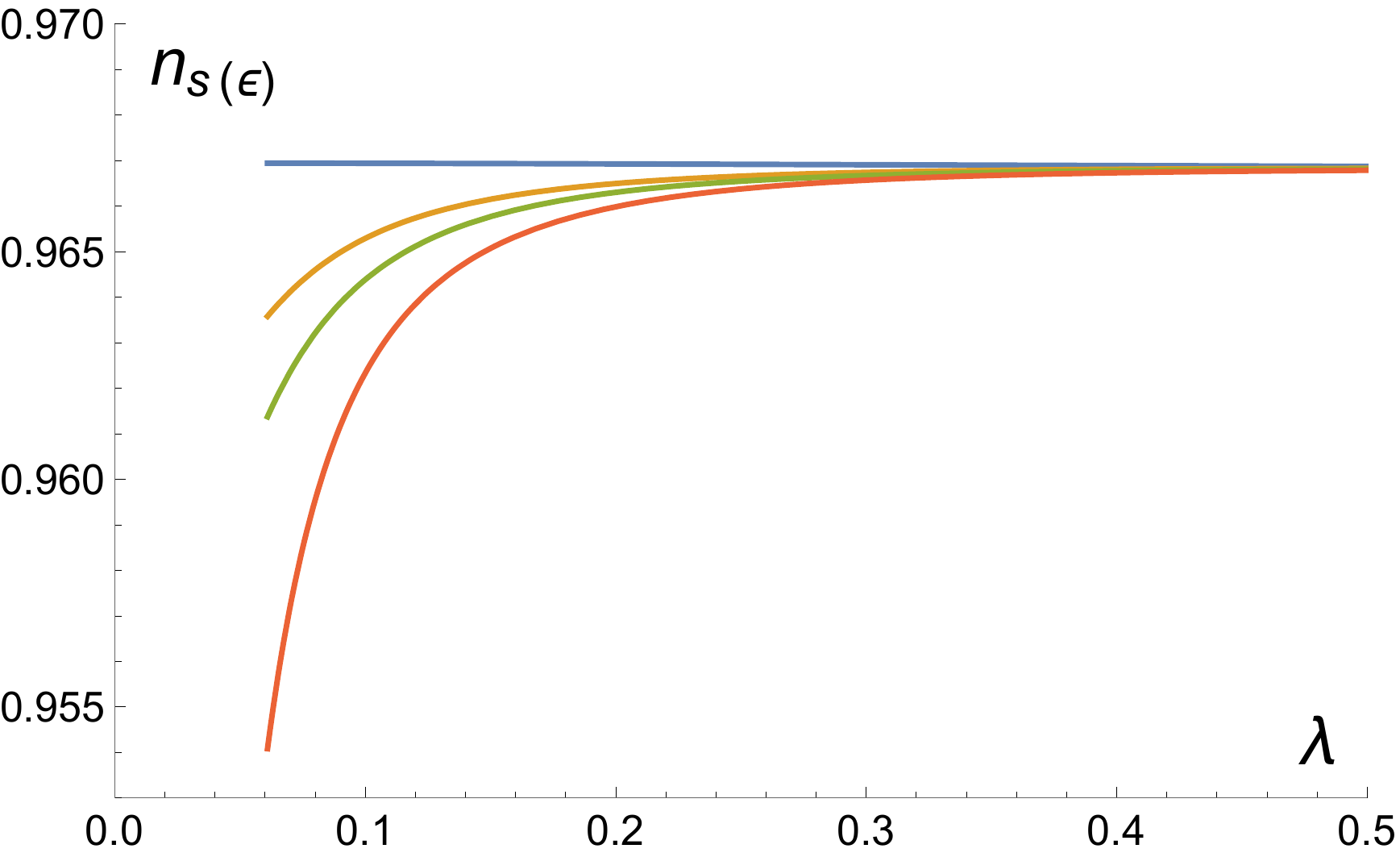}
\includegraphics[trim = 0mm  0mm 1mm 1mm, clip, width=7.5cm, height=5.5cm]{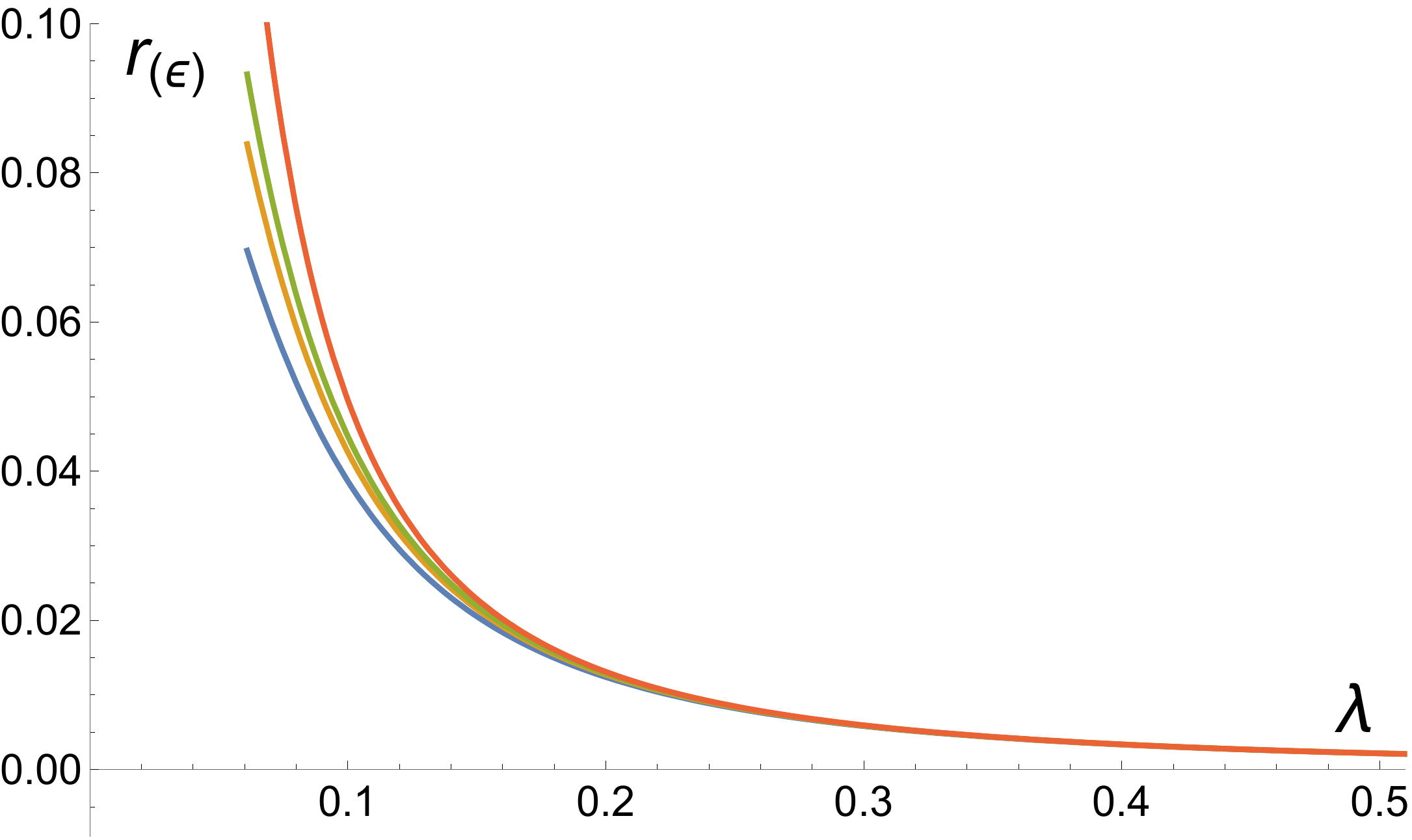}
\caption{\small The l.h.s. figure illustrates the scalar spectral index $n_s$ as a function of $\lambda$ as given by Eq.~\eqref{ns2a} for $N_{ke}=60$ and (from top to bottom) $p=2,3,4,10^6$ while the r.h.s. figure shows $r_{(\epsilon)}(\lambda)$ given by Eq.~\eqref{r2b}, also for $N_{ke}=60$ and (from left to right) $p=2,3,4,10^6$, both for $\lambda<\lambda_l$. The subindex $\epsilon$ is intended to emphasize that the solution has been obtained under the condition $\epsilon=1$ for ending inflation.
}
\label{nseps}
\end{center}
\end{figure}

The number of e-folds Eq.~\eqref{Nke} is
\begin{equation}
N_{ke\epsilon}(n_s,\lambda,p) = \frac{4p\lambda^2-\delta_{n_s}R_2+\sqrt{\delta_{n_s}^2+4p^2\lambda^2\delta_{n_s}+16p^2\lambda^4}}{4p\lambda^2\delta_{n_s}}\;,
\label{Nke2a}
\end{equation}
where $R_2=\sqrt{1+2p^2\lambda^2}$ and, in analogy with  Eq.~\eqref{Nke1a},  the subindex $\epsilon$ appears here to remind us that $N_{ke}$ is obtained from the solution to the $\epsilon=1$ condition but then it is dropped from the following expressions. From Eq.~\eqref{Nke2a} we can solve for $n_s$ (see Fig.\,\ref{nseps})
\begin{equation}
n_s=\frac{1+4p^2\lambda^2-(R_2+4(-2+N_{ke})p\lambda^2)(R_2+4N_{ke}p\lambda^2)}{1-(R_2+4N_{ke}p\lambda^2)^2}\;,
\label{ns2a}
\end{equation}
for large $N_{ke}$
\begin{equation}
n_s=1-\frac{2}{N_{ke}}+\frac{2R_2-p}{4p\lambda^2N_{ke}^2}+\cdot\cdot\cdot\;.
\label{ns2abexp}
\end{equation}

To obtain $N_{ke}$ in terms of $r$ we proceed in a similar way as in Eq.~\eqref{r1b}. From Eqs.~\eqref{Nke}, \eqref{fik1b} and \eqref{fie2a} we get
\begin{equation}
N_{ke\epsilon}(r,\lambda,p) = \frac{\sqrt{r+32p^2\lambda^2}-\sqrt{r}\sqrt{1+2p^2\lambda^2}}{4p\lambda^2\sqrt{r}}\;,
\label{Nke2b}
\end{equation}
solving for $r$ (see Fig.\,\ref{nseps})
\begin{equation}
r=\frac{16p}{p+8N_{ke}^2p\lambda^2+4N_{ke}\sqrt{1+2p^2\lambda^2}}\;,
\label{r2b}
\end{equation}
with large $N_{ke}$ expansion
\begin{equation}
r=\frac{2}{\lambda^2N_{ke}^2}-\frac{\sqrt{1+2p^2\lambda^2}}{p\lambda^4N_{ke}^3}+\cdot\cdot\cdot\;.
\label{r2bexp}
\end{equation}
The previous results for $n_s$ in the large $N_{ke}$ expansion Eqs.~\eqref{ns1aexp} and \eqref{ns2abexp} give a leading term $n_s=1-\frac{2}{N_{ke}}+\cdot\cdot\cdot $. Also Eqs.~\eqref{r1bexp} and \eqref{r2bexp} give a leading term $r=\frac{2}{\lambda^2N_{ke}^2}+\cdot\cdot\cdot$ for the large $N_{ke}$ expansion of $r$.
\section {\bf Removing the $\lambda$ and $V_0$ parameters }\label{REM}
We can express the parameter $\lambda$ purely in terms of $p$ and the observables $n_s$ and $r$ by substituting e.g., $\phi_k$ from Eq.~\eqref{fik1a} in Eq.~\eqref{Int}, $r=16\epsilon$, and solving for $\lambda$ 
\begin{equation}
\lambda=\frac{\sqrt{p^2(8\delta_{n_s}-r)^2-4r^2}}
{8p\sqrt{2r}}\;.
\label{lambdaeq}
\end{equation}
Technically, the parameter $\lambda$ can take values from 0 to $\infty$. The limiting value $\lambda=0$ occurs for  $r=\frac{8p}{p+2}\delta_{n_s}$ or, equivalently, $n_s=1-\frac{p+2}{8p}r$ which is exactly the relation between $n_s$ and $r$ for monomial potentials of the form $V(\phi)\sim \phi^p$. Clearly $\lambda \rightarrow \infty$ when $r\rightarrow 0$.
\begin{figure}[tb]
\begin{center}
\includegraphics[width=10cm]{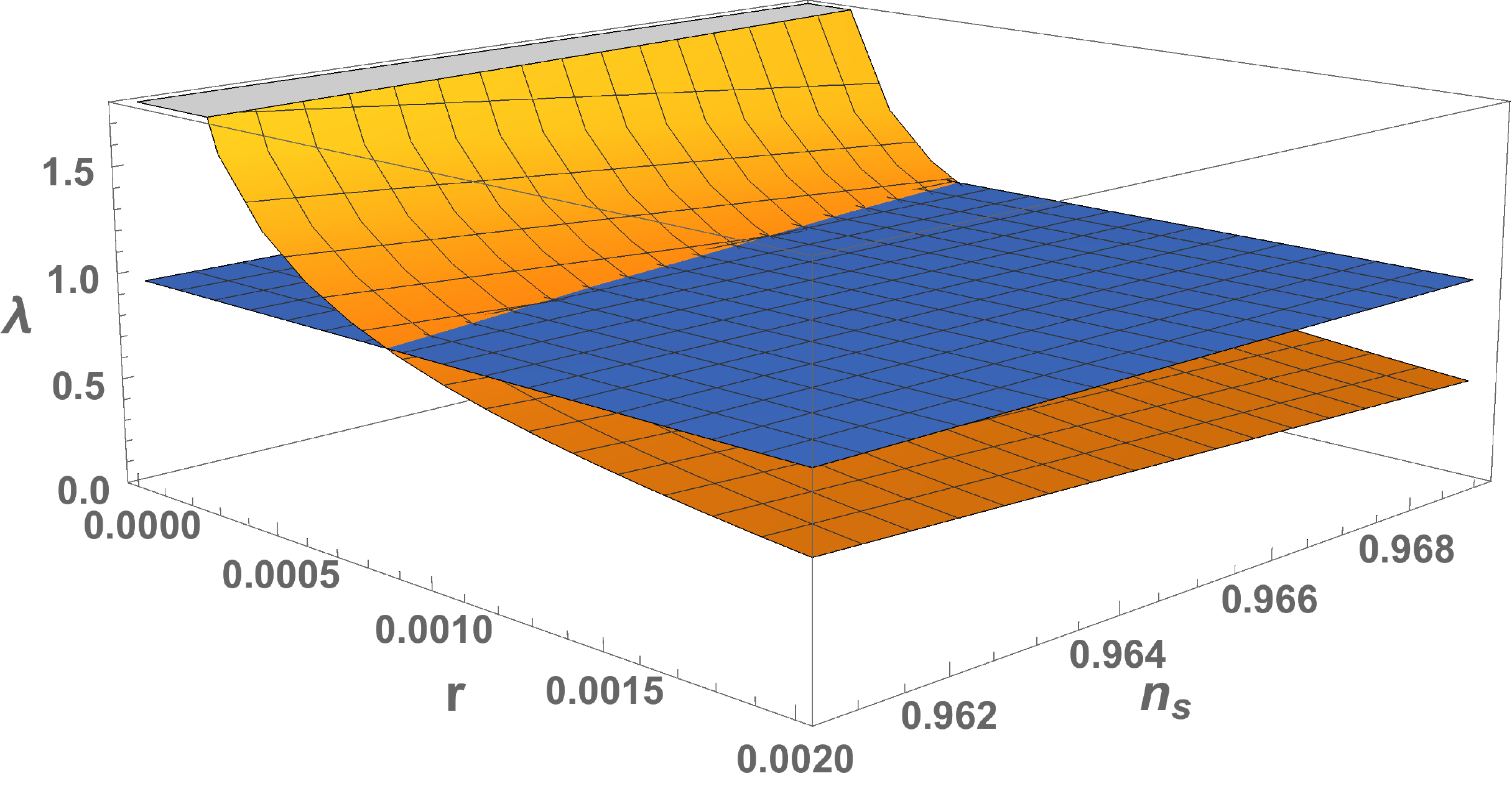}
\caption{\small Plot of the parameter $\lambda$ as given by Eq.~\eqref{lambdaeq} for $p=2$ as a function of $n_s$ and $r$. The horizontal plane shows the limiting value $\lambda_l$ of Eq.~\eqref{Lmin} which separates the parameter region where $\eta=-1$ ends slow-roll ($\lambda>\lambda_l$) from the region where inflation is terminated by the $\epsilon=1$ condition ($\lambda<\lambda_l$). These conditions on $\lambda$ are translated in conditions on $r$ for a fixed value of $n_s$ (see Eq.~\eqref{rl}).
}
\label{lam}
\end{center}
\end{figure}

The limiting value $\lambda_l$, defined by Eq.~\eqref{Lmin}, distinguishes between ending slow-roll by $\eta=-1$ or $\epsilon=1$. For $\lambda$ given by \eqref{lambdaeq} above, $\lambda_l$ translates into a limiting value for $r$, for a fixed value of $n_s$,  as follows
\begin{equation}
r_l=\frac{8p}{p^2-4}\left(4\sqrt{p^2-1}+p(4+\delta_{n_s})-2\sqrt{2p(4+\delta_{n_s})(p+\sqrt{p^2-1})+\delta_{n_s}^2-4}\right)\;,
\label{rl}
\end{equation}
thus, 
\begin{equation}
\lambda\geq\lambda_l \quad \Longleftrightarrow \quad r\leq r_l, \quad \lambda<\lambda_l \quad \Longleftrightarrow \quad r>r_l\;.
\label{lambdalrl}
\end{equation}
When $p=2$ we have
\begin{equation}
r_l=\frac{4\delta_{n_s}^2}{4+2\sqrt{3}+\delta_{n_s}}\;,
\label{rl1}
\end{equation}
the case $p=2$ implies $r_l=(6.7 \pm 1.6)\times 10^{-4}$ for $n_s=0.9649\pm0.0042$ \cite{Akrami:2018odb} . 
When the condition $\eta=-1$ is met $\lambda\geq\lambda_l$, and $r$ should be less than $r_l$ while values of $r$ larger than $r_l$ correspond to $\lambda<\lambda_l$  which occur when the condition $\epsilon=1$ ends inflation (see Fig.\,\ref{lam}) thus, the condition $\eta=-1$ is only satisfied for small values of $r$. For example, for  $p\geq1$ the largest value of $r_l$ is $r_l\approx 1.53\times 10^{-3}$ which occurs for $n_s=0.9607$ while for $p\geq2$ the largest value is $r_l\approx 8.23\times 10^{-4}$ also at $n_s=0.9607.$

For small $r$, $\lambda $ has the following expansion
\begin{equation}
\lambda=\frac{\delta_{n_s}}{\sqrt{2r}}-\frac{\sqrt{r}}{8\sqrt{2}}-\frac{r^{3/2}}{32\sqrt{2}p^2\delta_{n_s}}+\cdot\cdot\cdot\;,
\label{lnsrexp}
\end{equation}
thus, to leading order in $r$, $\lambda$ is $p$-independent. However, from the bounds $n_s=0.9649\pm 0.0042$ and $r<b$, the parameter $\lambda$ is bounded from below as $ \lambda > \frac{1}{16\sqrt{2b}}\sqrt{4(8(n_{s}^u-1)+b)^2-b^2}$  where $n_{s}^u$ is the upper bound for $n_s$. For the current bounds $n_{s}^u=0.9691$ and $b=0.063$, a lower bound for $\lambda$ is implied as $\lambda > 0.061$.
\noindent
\subsection {\bf General solutions for $r(n_s,N_{ke},p)$\label{SOL}}

\begin{figure}[tb]
\begin{center}
\includegraphics[trim = 0mm  0mm 1mm 1mm, clip, width=7.5cm, height=5.5cm]{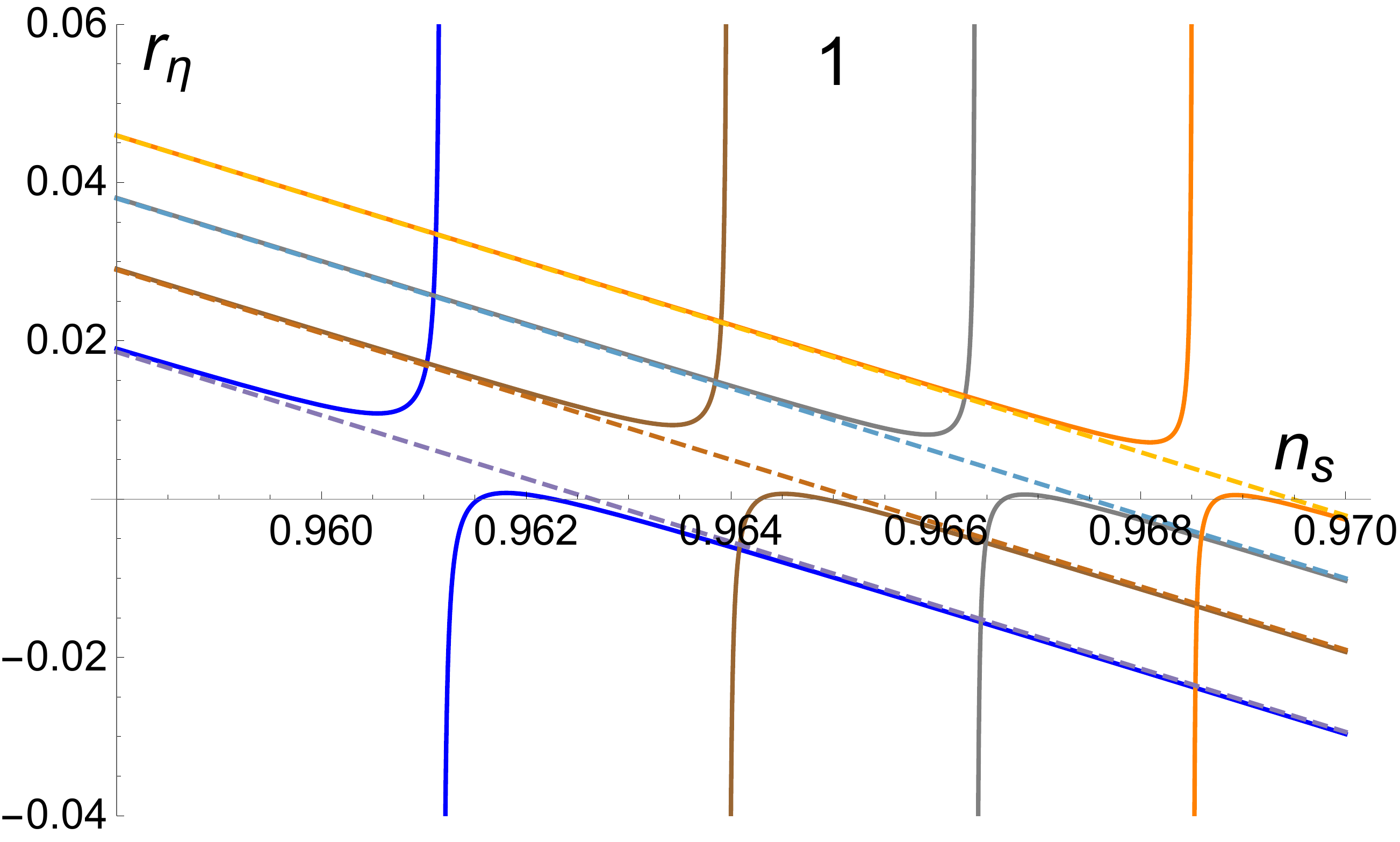}
\includegraphics[trim = 0mm  0mm 1mm 1mm, clip, width=7.5cm, height=5.5cm]{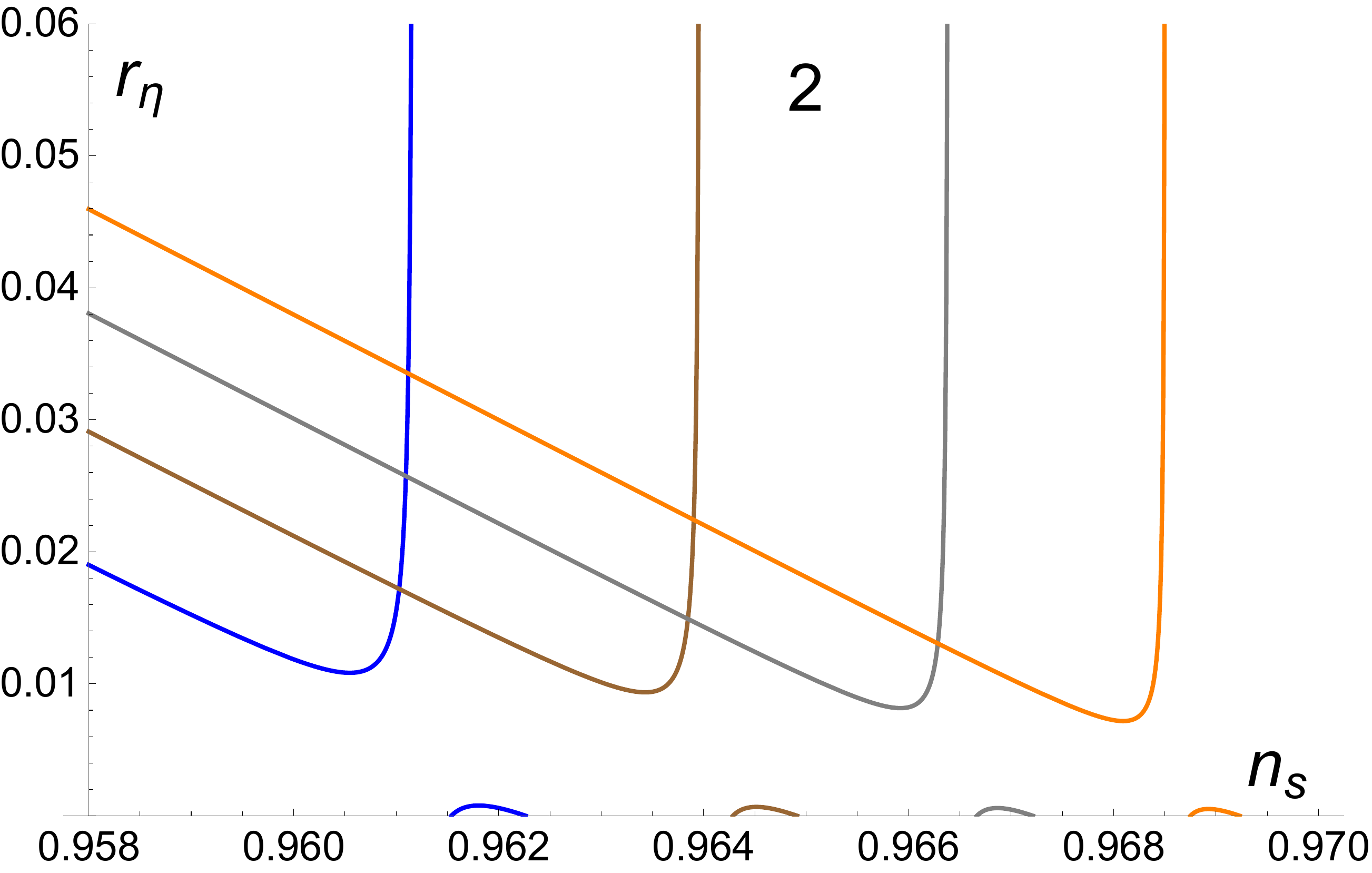}
\includegraphics[trim = 0mm  0mm 1mm 1mm, clip, width=7.5cm, height=5.5cm]{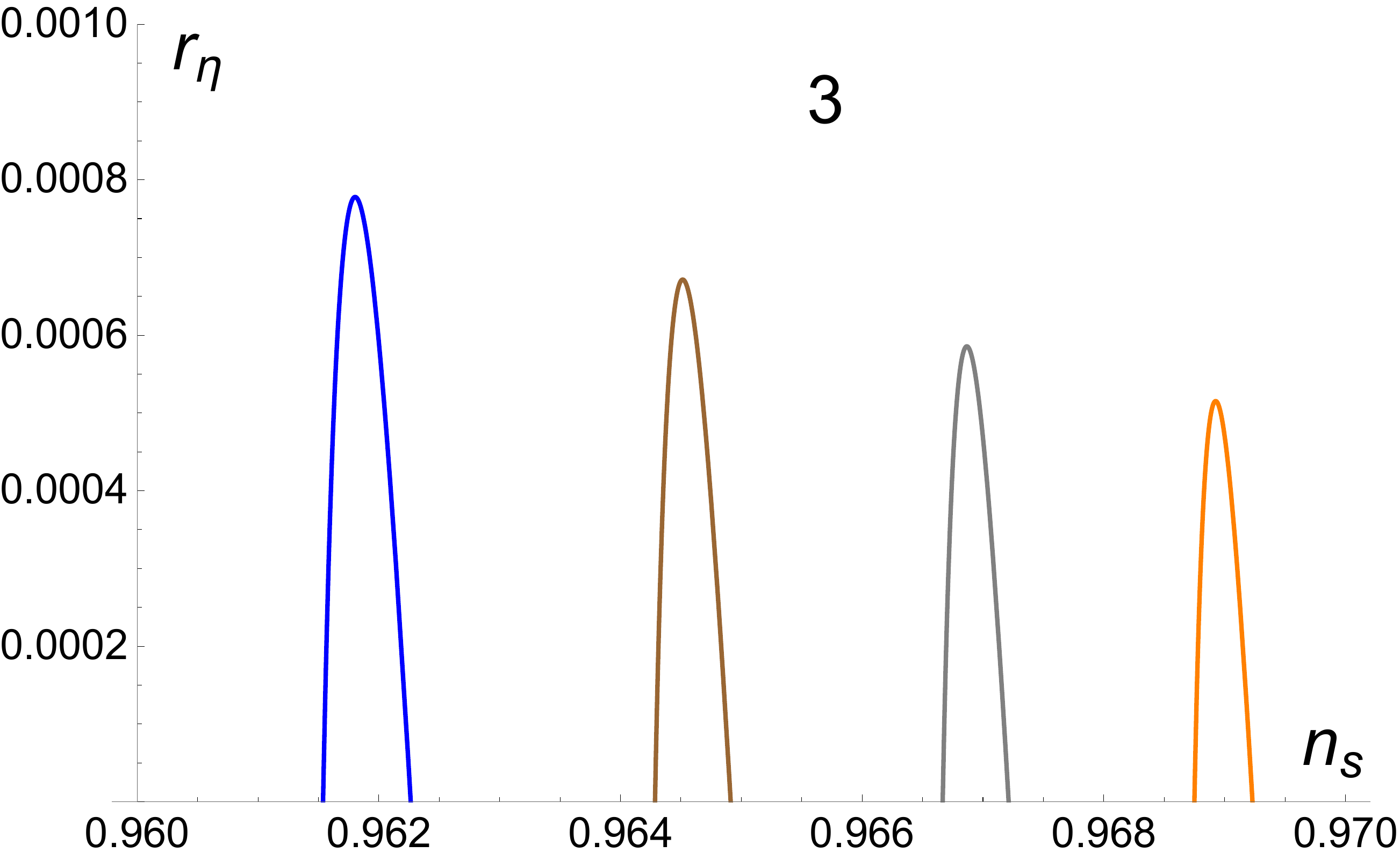}
\includegraphics[trim = 0mm  0mm 1mm 1mm, clip, width=7.5cm, height=5.5cm]{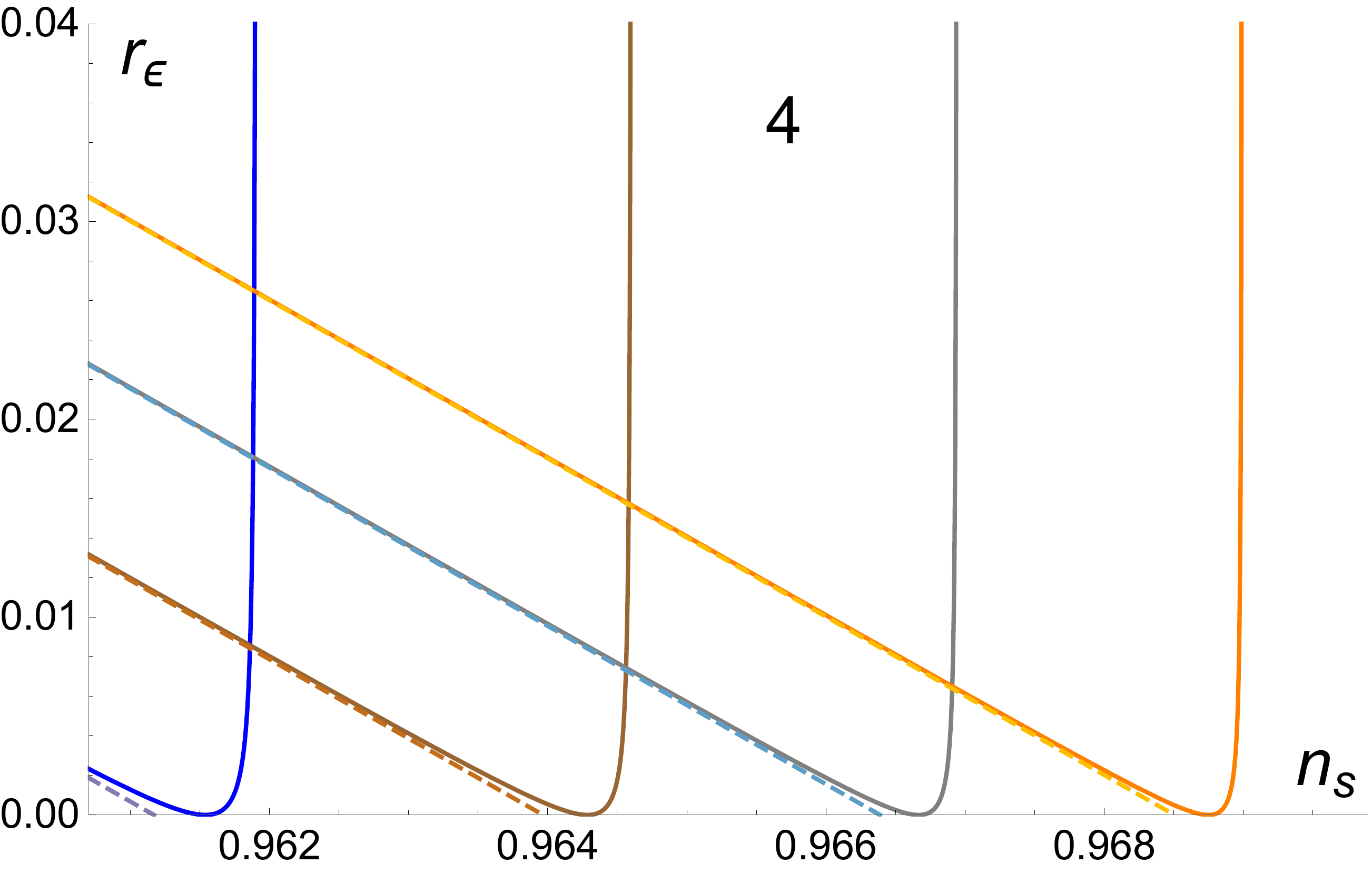}
\caption{\small In panel 1 we plot the solution for $r$ as a function of $n_s$ for $p=2$ and (from left to right) $N_{ke}=52$, 56, 60 and 64 (Eq.~\eqref{reta2(1)}). Dashed lines correspond to the large $N_{ke}$ approximation given by Eq.~\eqref{reta2(1)exp}. The negative part of the solutions as well as the approximations are discarded in panel 2. The very small bumps in panel 2 are the relevant solutions because $r_{\eta}$ is only valid for small $r$ (less than $\approx 0.00082$ for the $p=2$ case). This bumps are amplified in panel 3 and from them only the l.h.s. part of the curve is  the solution which matters, because it connects smoothly with the $r_{\epsilon}$ solution shown in panel 4. 
Because $r_{\epsilon}$ and its approximation as given by Eqs.~\eqref{reps2(1)} and \eqref{reps2(1)exp} are valid for larger values of $r$ ($\lambda<\lambda_l$) and because they are very close to the $r_{\eta}$ solution in the small $r$ regime (large $\lambda$) it is then posible to use the solution $r_{\epsilon}$ of \eqref{reps2(1)} for the whole range of $r$ with negligible error in the very small $r$ regime. Thus, the (finally) relevant solution can be considered as the part of $r_{\epsilon}$ in panel 4 which ascend almost vertically from $r=0$ to end in the monomials as shown in  Fig.\,\ref{figp2} for $p=2$, $N_{ke}=50$, 60 and for other values of $p$ in Fig.\,\ref{alfamono} where we plot Eq.~\eqref{reps1} .
}
\label{r(1)}
\end{center}
\end{figure}

\begin{figure}[tb]
\begin{center}
\includegraphics[trim = 0mm  0mm 1mm 1mm, clip, width=7.cm, height=6.cm]{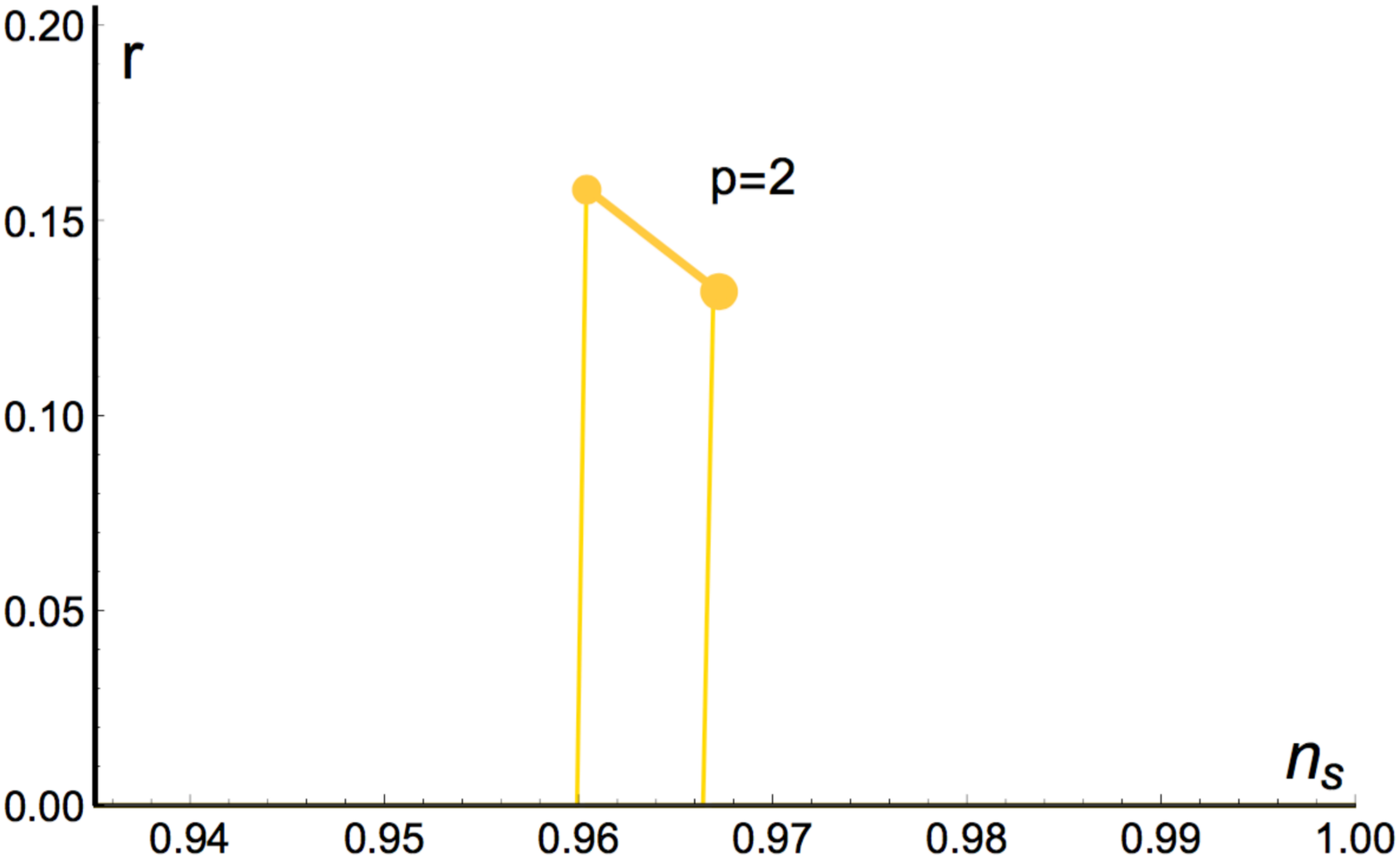}
\includegraphics[trim = 0mm  0mm 1mm 1mm, clip, width=8.cm, height=6.cm]{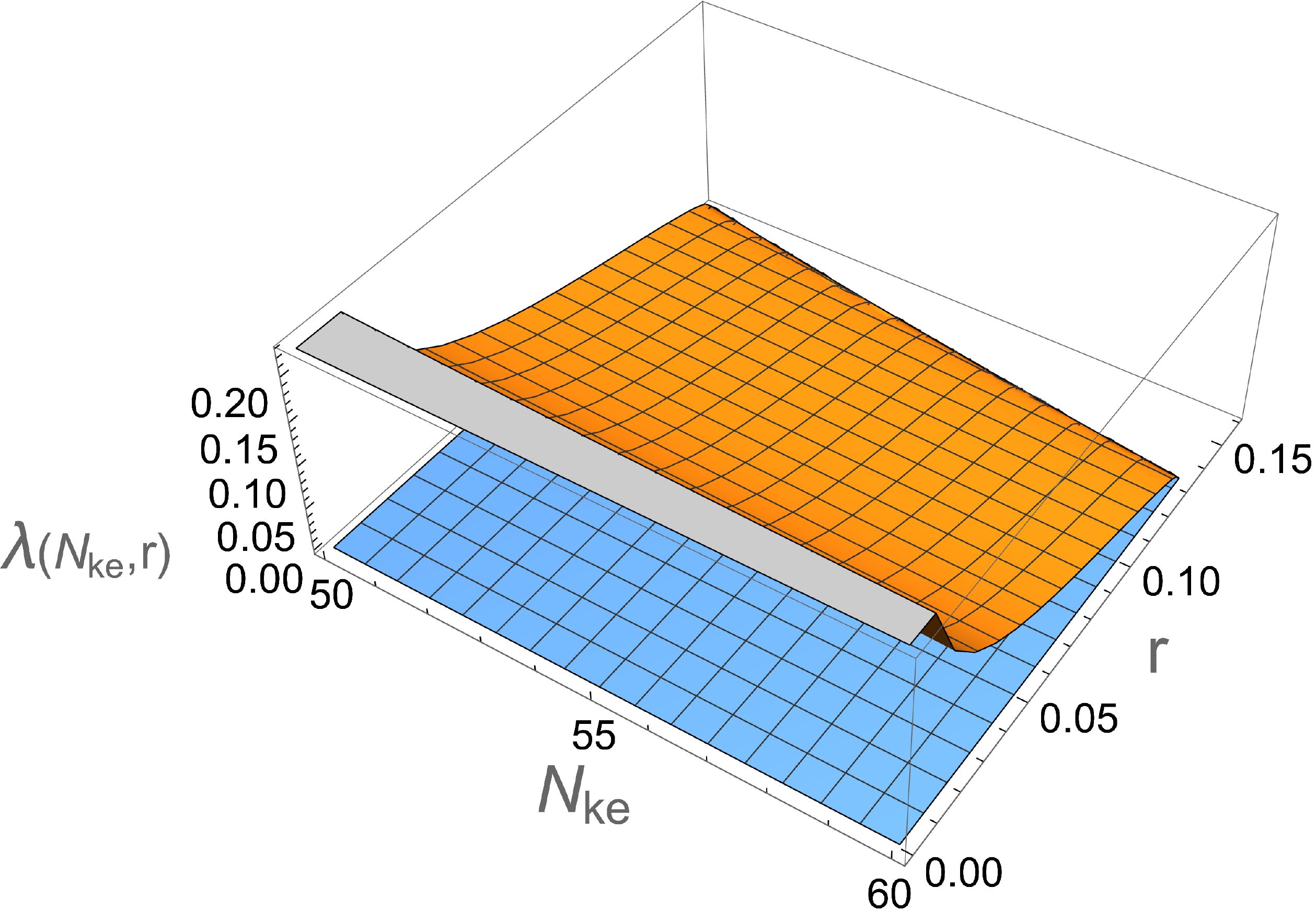}
\caption{\small In the l.h.s. figure we plot $r$ as a function of $n_s$ as given by  Eq.~\eqref{reps2(1)} for $p=2$, $N_{ke}=50$ (joining the small circle) and $N_{ke}=60$ reproducing the numerical calculation shown in the Fig.~8 of the Planck 2018 Collaboration \cite{Akrami:2018odb}  and duplicated here below as Fig.\,\ref{Plancky}. The r.h.s. figure is a similar plot with a trivial extension in the $N_{ke}$ direction. The horizontal plane is just for reference to the $\lambda=0$ limiting value. 
}
\label{figp2}
\end{center}
\end{figure}

We can eliminate the parameter $\lambda$ in all the previous equations in such a way that only the observables $n_s$ and $r$ appear. First we separately discuss the $p=2$ case by substituting $\lambda$ from Eq.~\eqref{lambdaeq} into the expressions for $N_{ke}$ given by Eqs.~\eqref{Nke1a}, \eqref{Nke1b}, \eqref{Nke2a}, \eqref{Nke2b} and solving for $r$ obtaining the following $two$ independent solutions: one for the case $\lambda\geq \lambda_l$ where the end of slow-roll is given by the condition $\eta=-1$ and labeled by the symbol $\eta$  and one for the case $\lambda< \lambda_l$ where the end of inflation is given by the condition $\epsilon=1$, labeled by the symbol $\epsilon$
\begin{equation}
r_{\eta }(p=2) =\frac{4(N_{ke}\delta_{n_s}-2)((N_{ke}+1)\delta_{n_s}-2)}{N_{ke}((N_{ke}+1)\delta_{n_s}-2)-3},
\label{reta2(1)}
\end{equation}
\begin{equation}
r_{\epsilon}(p=2) =\frac{4(N_{ke}\delta_{n_s}-2)^2}{N_{ke}(N_{ke}\delta_{n_s}-2)+1}.
\label{reps2(1)}
\end{equation}
The approximations for large $N_{ke}$ are (see Figs.\,\ref{r(1)} and \ref{figp2} for the $p=2$ case).
\begin{equation}
r_{\eta }(p=2) =\frac{4(N_{ke}\delta_{n_s}-2)}{N_{ke}}+\frac{12}{N_{ke}^2}+\cdot\cdot\cdot\,,
\label{reta2(1)exp}
\end{equation}
\begin{equation}
r_{\epsilon}(p=2) =\frac{4(N_{ke}\delta_{n_s}-2)}{N_{ke}}-\frac{4}{N_{ke}^2}+\cdot\cdot\cdot\,.
\label{reps2(1)exp}
\end{equation}
\begin{figure}[tb]
\begin{center}
\includegraphics[width=12cm]{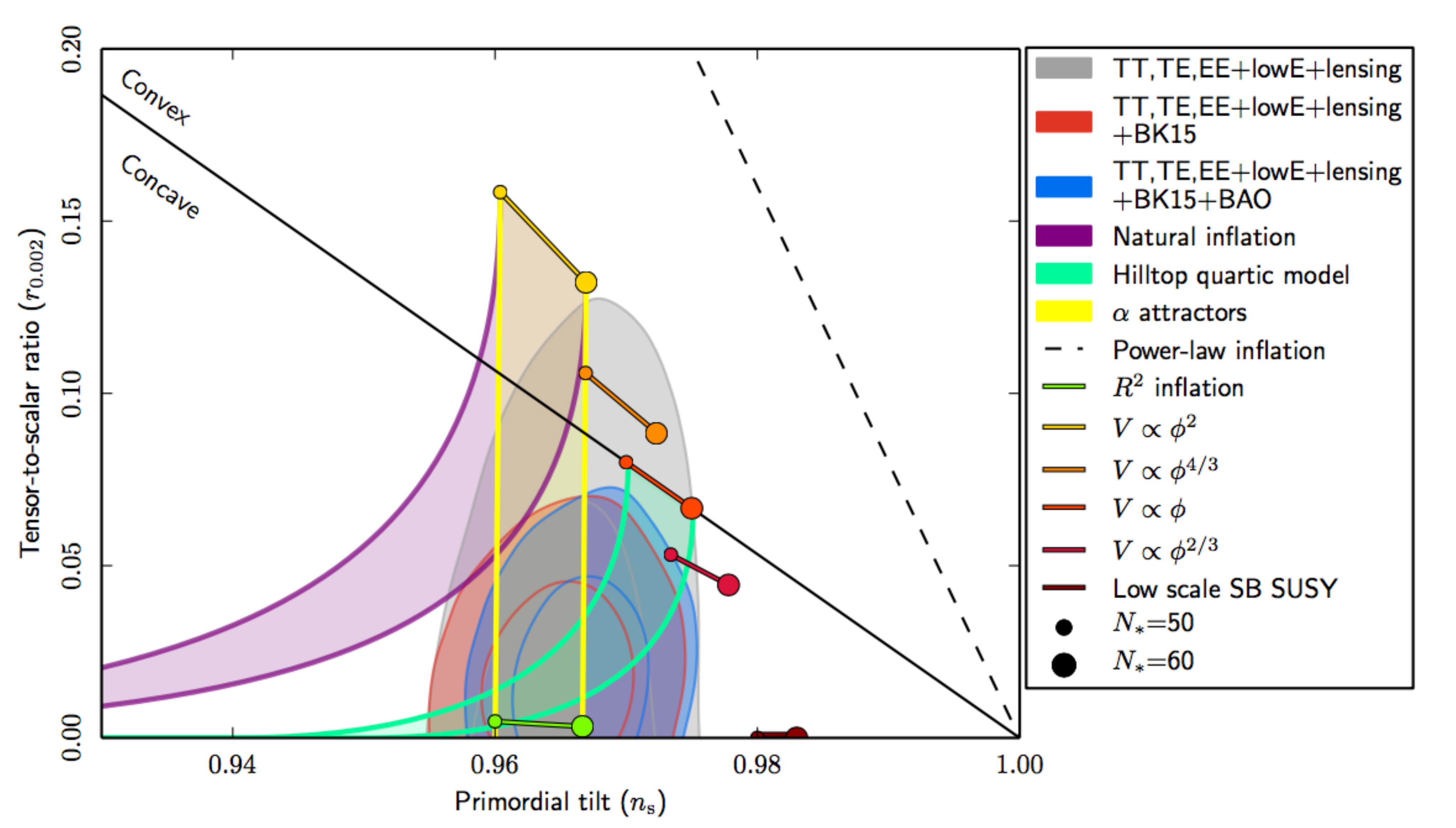}
\caption{\small We take figure 8 of the Planck Collaboration 2018 article \cite{Akrami:2018odb}  where monomial potentials are also considered together with several other interesting models of inflation (see description in the right hand side panel of the figure). From the figure we see that there is a substantial overlap of the predictions of $\alpha$-attractor inflation (yellow, almost vertical curves for the $p=2$ case) with Planck alone and in combination with BICEP2/Keck Array (BK15) \cite{BICEP2:2015ns} or BICEP2/Keck Array+Baryon Acoustic Oscillations (BK15+BAO) data. For other values of $p$ compare with the plot of Eq.~\eqref{reps1} in Fig.\,\ref{alfamono} below.
}
\label{Plancky}
\end{center}
\end{figure}
\begin{figure}[tb]
\begin{center}
\includegraphics[width=12cm]{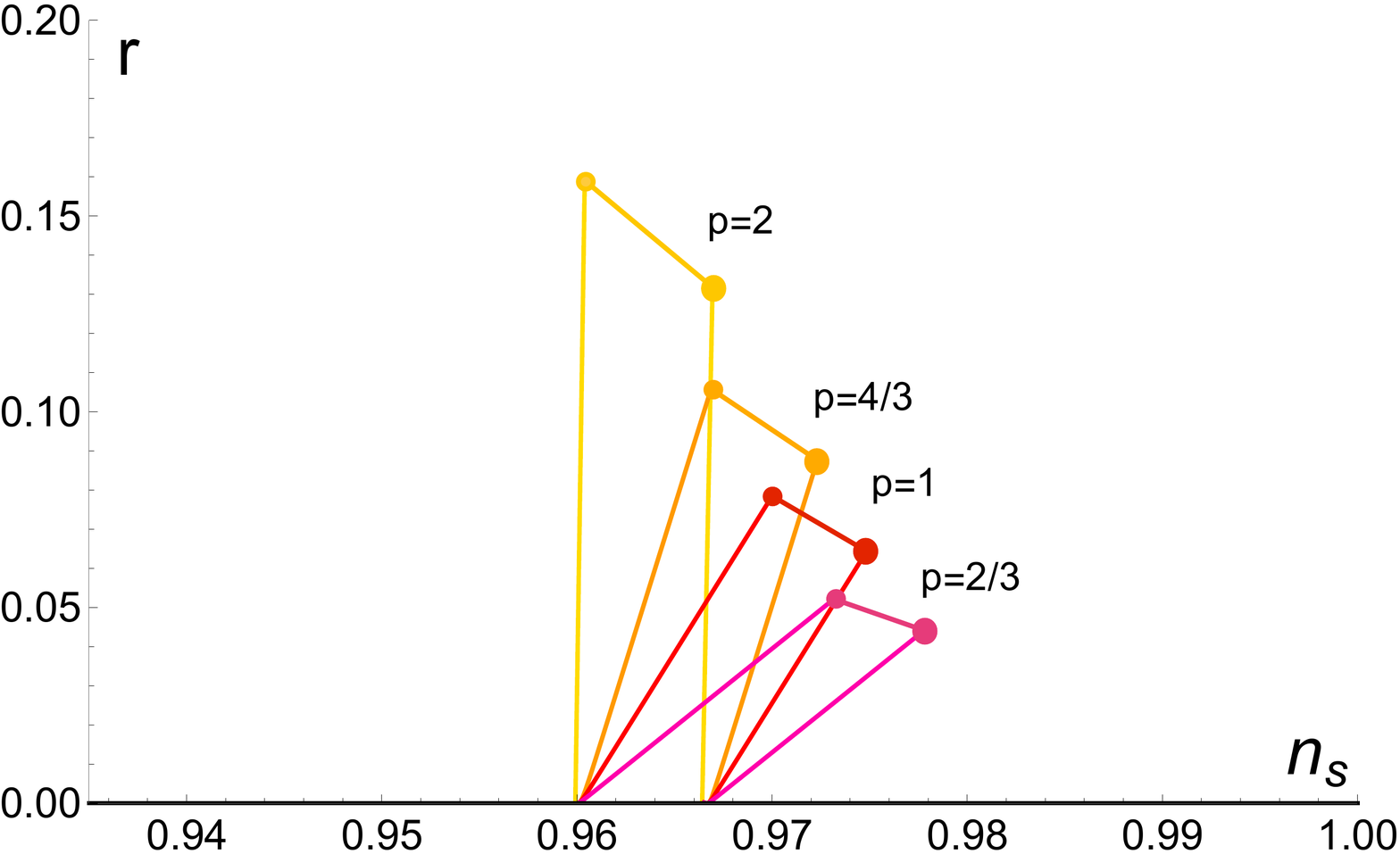}
\caption{\small Plot of $r_{\epsilon}$ as given by Eqs.~\eqref{reps2(1)} and \eqref{reps1} as a function of $n_s$ for $N_{ke}=50$ (joining the small circles) and $N_{ke}=60$ for $p=2$, 4/3, 1 and 2/3 as shown, appearing in the Fig.~8 of the Planck 2018 Collaboration and duplicated here as Fig.\,\ref{Plancky} above. We see how the $\alpha$-attractor solutions practically cover the entire region of Planck alone and in combination with BK15 or BK15+BAO data.
}
\label{alfamono}
\end{center}
\end{figure}
\noindent 
For $p\neq 2$ we can proceed in an analogous way as above and solving in each case for $r$ in terms of $p$, $N_{ke}$ and $\delta_{n_s}$ obtaining the following two independent solutions
\begin{equation}
r_{\eta} =\frac{8p}{N_{ke}(N_{ke}+1)(p^2-4)}\left( p(N_{ke}((N_{ke}+1)\delta_{n_s}-2)-3)+R_3\right ),
\label{reta2}
\end{equation}
where $R_3=\left(9p^2+4N_{ke}((N_{ke}+1)\delta_{n_s}-2)(-2-p^2+N_{ke}((N_{ke}+1)\delta_{n_s}-2)\right)^{1/2}$ and 
\begin{equation}
r_{\epsilon} =\frac{8p}{N_{ke}^2(p^2-4)}\left(p(N_{ke}(N_{ke}\delta_{n_s}-2)+1)-R_4\right),
\label{reps1}
\end{equation}
where $R_4=\left(p^2+2N_{ke}(N_{ke}\delta_{n_s}-2)(2N_{ke}(N_{ke}\delta_{n_s}-2)+p^2)\right)^{1/2}$. Approximations for large $N_{ke}$ are 
as follows
\begin{equation}
r_{\eta} =\frac{8p}{p-2}\frac{(N_{ke}\delta_{n_s}-2)}{N_{ke}}-\frac{8p(p-1)}{p-2}\frac{1}{N_{ke}^2}+\cdot\cdot\cdot\,,
\label{reta2exp}
\end{equation}
\begin{equation}
r_{\epsilon} =\frac{8p}{p-2}\frac{(N_{ke}\delta_{n_s}-2)}{N_{ke}}-\frac{4p^2}{p+2}\frac{1}{N_{ke}^2}+\cdot\cdot\cdot\,,
\label{reps1exp}
\end{equation}
We expect the first term to dominate and should be positive thus, for $p>2$, $n_s<1-2/N_{ke}$ while $p<2$ requires $n_s>1-2/N_{ke}$ or, perhaps more appropriate, for $p>2$, $N_{ke}>\delta_{n_s}/2$ while $p<2$ requires $N_{ke}<\delta_{n_s}/2$. From the Planck bounds $n_s=0.9649\pm0.0042$ it follows that $50.9<N_{ke}<64.7$ for any $p$. For a dominant first term we also expect that $r<b$ where $b=0.063$ is the upper bound for $r$ reported by the Planck 2018 Collaboration \cite{Akrami:2018odb} . Thus assuming that 
\begin{equation}
r \approx \frac{8p}{p-2}\frac{(N_{ke}\delta_{n_s}-2)}{N_{ke}} < b\,,
\label{rbound}
\end{equation}
we get
\begin{equation}
p >  \frac{2 b N_{ke}^l}{16+N_{ke}^l(b-8(1-n_s^u))}\approx 0.97\,,
\label{pbound}
\end{equation}
where $N_{ke}^l\approx 50.9$ is the lower bound for $N_{ke}$ and $n_s^u\approx 0.9691$ is the upper bound for $n_s$. The tensor-to-scalar ratio $r$ is determined at $\phi_k$ thus, we conclude that $p$ should be bigger than $\approx 0.97$ at the scale of wavenumber mode $k$.

For the $p=2$ case, the bound $\lambda\geq \lambda_l= (2+\sqrt{3})^{1/2}/2  \approx 0.9659$  is equivalent to a bound on $r$ as $r < 0.00082$ for $n_s=0.9607$. Thus, the expression for $r_{\eta}$ is valid for very small $r$ only. In this regime the solution $r_{\epsilon}$ (which should be used for $\lambda<\lambda_l$ or $r>r_l \approx 0.00051$ at $n_s=0.9691$) is very close to the $r_{\eta}$ solution being possible to use it also for the very small $r$ regime (large $\lambda$ regime) with negligible error. The case $p\neq 2$ is not very different thus, in what follows, we study only the $r_{\epsilon}$ solution for all possible values of $r$.
In Fig.\,\ref{r(1)} we show the solutions \eqref{reta2(1)} and \eqref{reps2(1)} for the $p=2$ case reproducing in Fig.\,\ref{figp2} the numerical solution which appear in Fig.~8  of the Planck 2018 Collaboration \cite{Akrami:2018odb} (shown here as Fig.\,\ref{Plancky}).
This solution contains the monomial solution for $r(n_s)$ of the $\phi^2$ model. In Fig.\,\ref{alfamono} we plot several other cases of Eq.~\eqref{reps1} by giving $p$ the values $p=2, 4/3, 1$ and $2/3$ containing all the monomial solutions shown in Fig.\,\ref{Plancky}. These solutions are contained exactly in the $\alpha$-attractor models as shown analytically in the following subsection.
\subsection {\bf Monomials as particular cases of $\alpha$-attractors}\label{Mono}

For monomials of the form $V_{mon}=\frac{1}{2}m^{4-p}\phi^p$ we get $\phi_k=\frac{2\sqrt{2}\,p}{\sqrt{r}}$ and $\phi_e=\frac{p}{\sqrt{2}}$ from where it follows that
\begin{equation}
 r=\frac{16p}{4N_{ke}+p}\,,
\label{rmon}
\end{equation}
also
\begin{equation}
\delta_{n_s}=\frac{p+2}{8p}r\,.
\label{nsmon}
\end{equation}
Eqs.~\eqref{rmon} and \eqref{nsmon} follow from the condition $\epsilon=1$, this condition ends inflation whenever $2/3\leq p\leq2$. Por $p$ outside this range slow-roll is terminated by $\eta=-1$ for $p<2/3$ in which case $ r=8p/(2N_{ke}+p-p^2)$ or $\eta=+1$ for $p>2$ in this case $r=8p/(2N_{ke}+p-1)$. In any case for large $N_{ke}$ and $p<<N_{ke}$ they all have the same limit $\sim 4p/N_{ke}$.
Thus for $p=2$ and $N_{ke}=50$, $(n_s,r)=(\frac{97}{101},\frac{16}{101})\approx (0.9604,0.1584)$ and so on. In this way we can calculate all the points appearing as circles in Fig.\,\ref{Plancky}. The point $(0.9604,0.1584)$, for example, is reached by the solution $r_{\epsilon}(p=2)$ Eq.~\eqref{reps2(1)} for $N_{ke}=50$ as shown by the left line in Fig.\,\ref{figp2} and similarly for the $N_{ke}=60$ line on the r.h.s. For $p\neq 2$ Eq.~\eqref{reps1} applies. In this way we draw Fig.\,\ref{alfamono} connecting the $\alpha$-attractor T-models with all the monomial as shown. Analytically, we can see this  as follows:
for the $p=2$ case we sustitute Eq.~\eqref{nsmon}  in Eq.~\eqref{reps2(1)} and we get $r_{\epsilon}(p=2)=16/(2N_{ke}+1)$ which is Eq.~\eqref{rmon} (with $p=2$). In general (for $p\neq 2$) substituting Eq.~\eqref{nsmon} in Eq.~\eqref{reps1} gives 
\begin{equation}
r_{\epsilon}=\frac{16p}{4N_{ke}+p},
\label{rQ}
\end{equation}
which is Eq.~\eqref{rmon}, exactly. Thus, all the monomial models are contained as particular cases by the $\alpha$-attractor T-models. This is illustrated by Fig.\,\ref{alfamono} only for the monomials shown in Fig.~8 (reproduced here as Fig.\,\ref{Plancky}) of the Planck 2018 Collaboration \cite{Akrami:2018odb} . Even more to the point: {\it the monomials are the ending points of the attractors} \cite{Kallosh:2013yoa}. This can be seen as follows: if we substitute Eq.~\eqref{nsmon} in the equation defining $\lambda$ (Eq.~ \eqref{lambdaeq}) we find that $\lambda=0$ becoming imaginary after this point. To have a clear understanding of this phenomenon let us substitute equation  \eqref{lambdaeq} for $\lambda$ in Eq.~\eqref{pot}; the resulting potential can then be written as
\begin{equation}
 V=\frac{3}{2}A_s\pi^2r\left(\frac{\sqrt{p(8\delta_{n_s}-r)+2r}}{\sqrt{p(8\delta_{n_s}-r)-2r}}\right)^p\tanh^p\left(\frac{\sqrt{p^2(8\delta_{n_s}-r)^2-4r^2}}{8p\sqrt{2r}}\phi\right),
\label{potlam}
\end{equation}
where $V_0$ has been calculated from Eq.~\eqref{IA}. If we study the potential \eqref{potlam} in the limit where $\delta_{n_s}\rightarrow\frac{p+2}{8p}r$ we find that it reduces to the following  expression  
\begin{equation}
 V_{mon}=\frac{3}{2}A_s\pi^2r\left(\frac{r}{8p^2}\right)^{p/2} \phi^p.
\label{potmon}
\end{equation}
\begin{figure}[tb]
\begin{center}
\includegraphics[width=11cm]{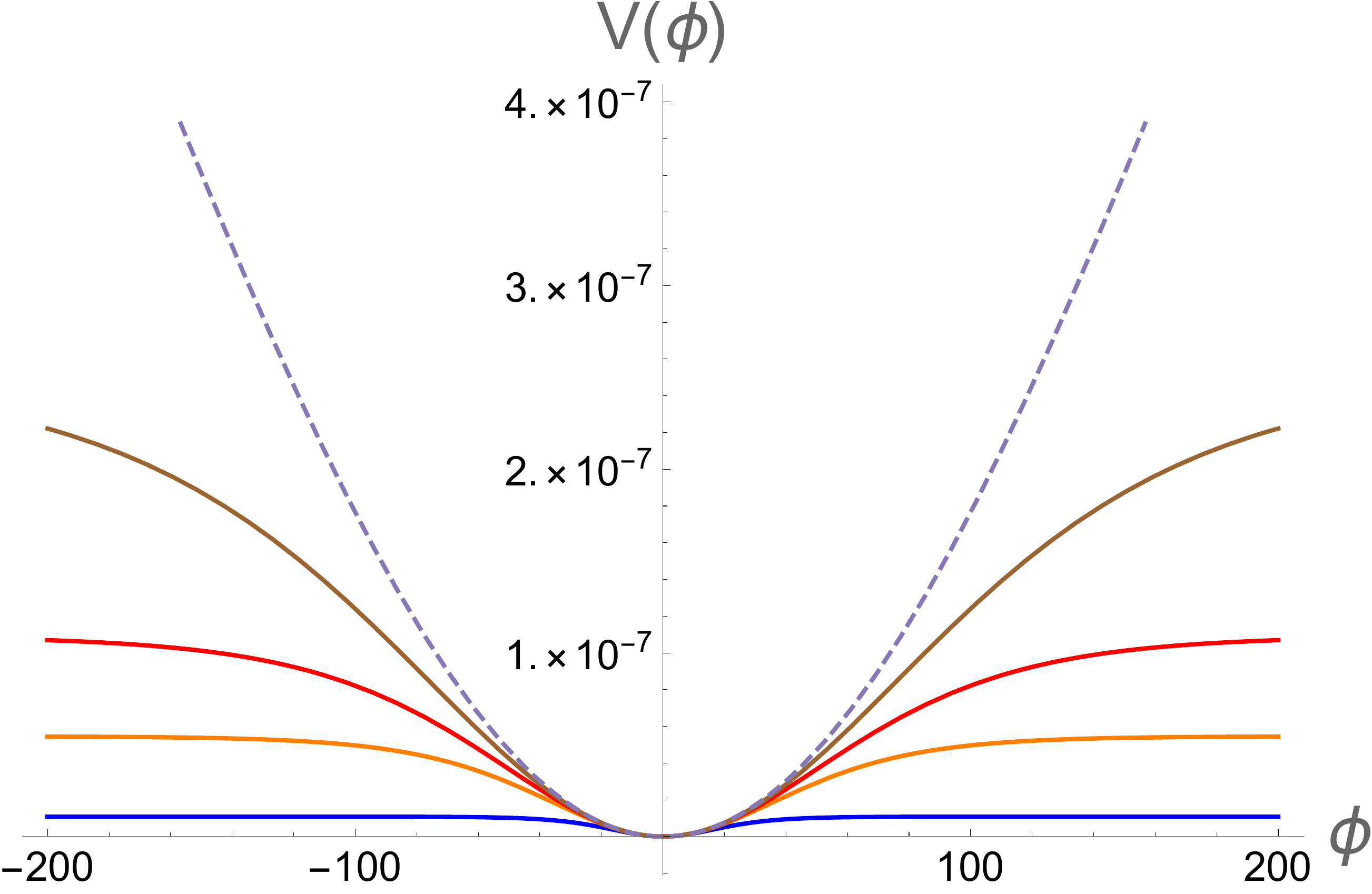}
\caption{\small Plot of the $\alpha$-attractor potential as given by Eq.~\eqref{potlam} as a function of $\phi$ for p=2 and $\delta_{n_s}=0.0351$ (equivalently $n_s=0.9649$) and various values of $r$ reaching $r=8p\delta_{n_s}/(p+2)$ (dashed curve) signaling the transition of the $\tanh^p(\lambda\phi)$ potential \eqref{potlam} to the monomial $\phi^p$ potential given by Eq.~\eqref{potmon}. The fact that the attractors end in monomials is now easily understood because the attractor potential transitions to the monomials potential when $\lambda=0$, that is, when $r=8p\delta_{n_s}/(p+2)$ which is precisely the relation between $r$ and $n_s$ for the monomials potential.
}
\label{pottanmon}
\end{center}
\end{figure}
One can easily check that the potential \eqref{potmon} corresponds exactly to monomials of the form $V_{mon}=\frac{1}{2}m^{4-p}\phi^p$. Thus, the fact that we reach the monomials in Fig.\,\ref{alfamono} is now easily understood because our original potential transitions to the monomials potential when $\lambda=0$, that is, when $r=8p\delta_{n_s}/(p+2)$ which is precisely the relation between $r$ and $n_s$  for the monomials potential.
The potential \eqref{potlam} is shown in Fig.\,\ref{pottanmon} as a function of $\phi$  for $p=2$ and the mean value $n_s=0.9649$ \cite{Akrami:2018odb} (or $\delta_{n_s}=0.0351$) for various values of $r$ reaching $r=8p\delta_{n_s}/(p+2)$ (dashed curve) signaling the transition of the $\tanh^p(\lambda\phi)$ potential \eqref{potlam} to the monomial $\phi^p$ potential of Eq.~\eqref{potmon}. The expresion $r=8p\delta_{n_s}/(p+2)$ makes $\lambda=0$ but $\lambda=0$ in what we saw before takes us to the ends of the curves in Fig.\,\ref{alfamono} reaching the predictions in the $n_s$-$r$ plane for the monomials potential. 
A similar study can be done with $\alpha$-attractor E-models defined by the potential \eqref{potEmodel}
which generalize the Starobinsky potential in the Einstein frame. In this case the generalized potential written in terms of $n_s$ and $r$
\begin{equation}
 V=\frac{3}{2}A_s\pi^2r\left(\frac{p(8\delta_{n_s}-r)}{8p\delta_{n_s}-(p+2)r}\right)^p\left(1-e^{-\frac{8p\delta_{n_s}-(p+2)r}{4p\sqrt{2r}}\phi}\right)^p\,,
\label{potstarob}
\end{equation}
also contains the monomials potential \eqref{potmon} much in the same way as discussed before (see Fig.\,\ref{potstaro}).
\begin{figure}[tb]
\begin{center}
\includegraphics[width=11cm]{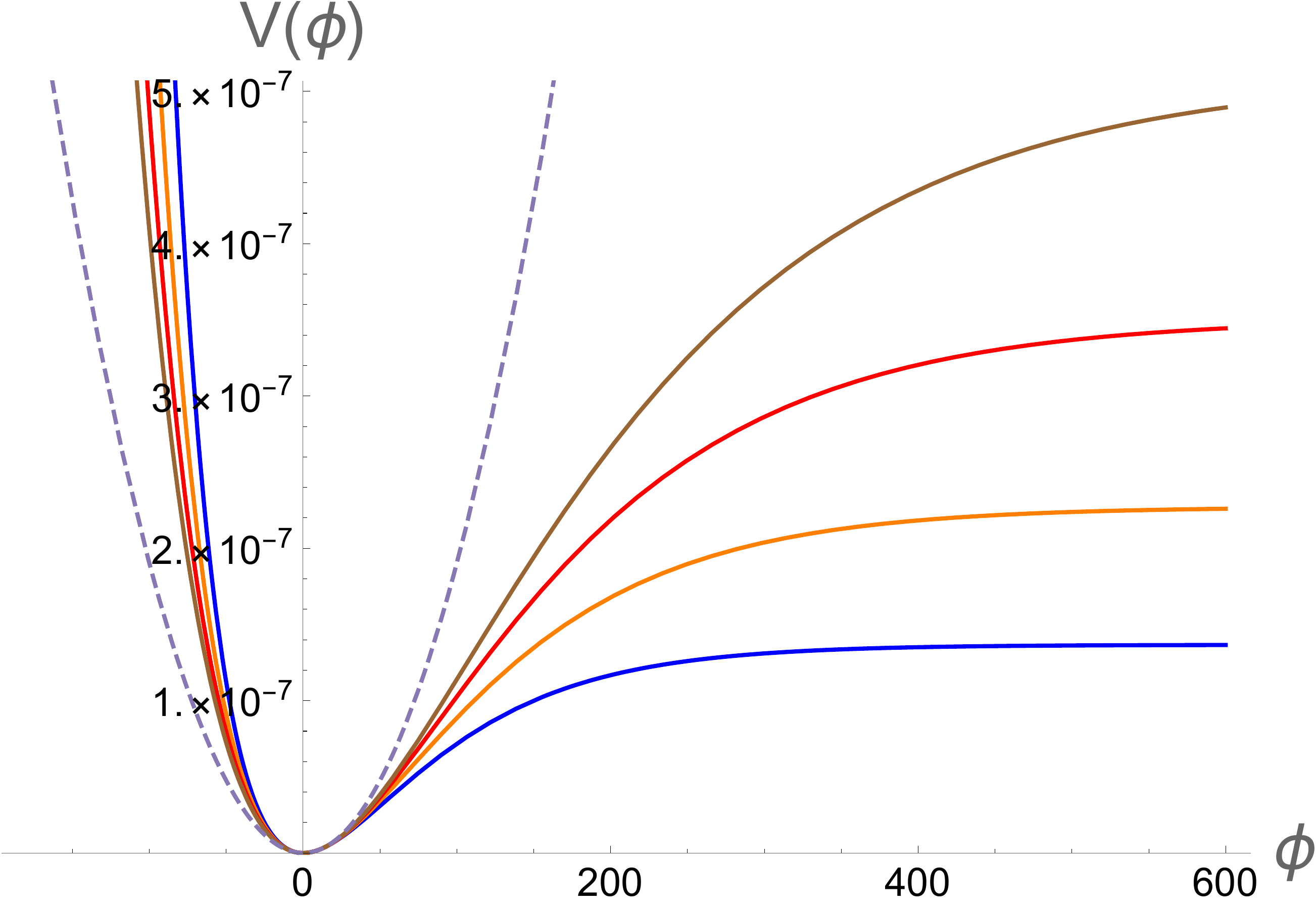}
\caption{\small As for the T-models the $\alpha$-attractor E-models also join and end in monomials much in the same way as T-models do. Here we show the potential as given by Eq.~\eqref{potstarob} as a function of $\phi$ for p=2, $\delta_{n_s}=0.0351$ (equivalently $n_s=0.9649$) and various values of $r$ reaching $r=8p\delta_{n_s}/(p+2)$ (dashed curve) signaling the transition of the  potential \eqref{potstarob} to the monomials $V_{mon}=\frac{1}{2}m^{4-p}\phi^p$ potential, given by Eq.~\eqref{potmon} in terms of the observable  $r$.
}
\label{potstaro}
\end{center}
\end{figure}
\section {\bf Conclusions}\label{CON}
We have carried out an analytical study of a class of $\alpha$-attractor T-models generalized from the simplest monomials and given by the potential $V = V_0\tanh^ p\left(\lambda \frac{\phi}{M_{pl}}\right)$ without paying attention to its origin, but dealing with it only as a phenomenological model of inflation. We can see how the analytical study clarifies several of its important properties and characteristics. In particular, we have obtained exact solutions, valid for any $p$, for the spectral index $n_s$ and for the tensor-to-scalar ratio $r$ in terms of the number of  e-folds $N_{ke}$ and the parameter $\lambda$. Eliminating the parameter $\lambda$ we can also obtain exact solutions for $r$ in terms of $n_s$ and $N_{ke}$. These solutions allow us to study the $n_s$-$r$ plane and compare with numerical studies such as those presented by the Planck Collaboration, reproducing their results. Our analytical study allows us to observe precisely how the monomial potentials are contained in the $\alpha$-attractor models and also constitute the end points of our solutions. Finally we show how in the appropriate limit the potential for the $\alpha$-attractor for both T and E-models exactly reduces to the monomials potential providing a clear explanation of the relationship between them.

\acknowledgments
I would like to thank the anonymous referee for a detailed and careful revision of the article and for useful advice. Financial support from UNAM-PAPIIT,  IN104119, {\it Estudios en gravitaci\'on y cosmolog\'ia} is gratefully acknowledged.

\end{document}